\documentclass[apj]{emulateapj}
\usepackage[dvips]{color}
\usepackage{graphicx}
\DeclareGraphicsExtensions{.pdf,.png,.jpg}

\newcommand*{\teff}{$T_{\rm eff}$}
\newcommand*{\logg}{$\log~g$}
\newcommand*{\feh}{[Fe/H]}
\newcommand*{\afe}{[$\alpha$/Fe]}
\newcommand*{\kms}{km s$^{-1}$}
\newcommand*{\zmax}{$Z_{\rm max}$}

\newcommand*{\vphi}{$V_{\rm \phi}$}

\newcommand*{\alp}{$\alpha$}

\usepackage[dvips]{color}
\usepackage{apjfonts}

\slugcomment{}

\shorttitle{CEMP Stars in the Halo Components of the Milky Way}
\shortauthors{Carollo et al.}

\begin{document}


\title{Carbon-Enhanced Metal-Poor Stars in the Inner and Outer Halo Components of the Milky Way}

\author{Daniela Carollo}
\affil{Research School of Astronomy and Astrophysics, Australian National University, Cotter Road, Weston, ACT 2611, Australia\\ 
INAF-Osservatorio Astronomico di Torino, Pino Torinese, Italy}
\email{carollo@mso.anu.edu.au}

\author{Timothy C. Beers}
\affil{Department of Physics \& Astronomy and JINA: Joint Institute for Nuclear Astrophysics, Michigan
State University,\\ E. Lansing, MI 48824, USA}
\email{beers@pa.msu.edu}

\author{Jo Bovy}
\affil{Center for Cosmology and Particle Physics
Department of Physics, New York University, 4 Washington Place, New York, NY 10003, USA}
\email{jb2777@nyu.edu}

\author{Thirupathi Sivarani}
\affil{Indian Institute of Astrophysics, II Block, Koramangala, Bangalore 560034, India}
\email{sivarani@iiap.res.in}

\author{John E. Norris, Ken C. Freeman}
\affil{Research School of Astronomy and Astrophysics, Australian National University, Cotter Road, Weston, ACT 2611, Australia}
\email{jen@mso.anu.edu.au; kcf@mso.anu.edu.au}

\author{Wako Aoki}
\affil{National Astronomical Observatory, Mitaka, Tokyo, 181-8588, Japan}
\email{aoki.wako@nao.ac.jp}

\author{Young Sun Lee, Catherine R. Kennedy}
\affil{Department of Physics \& Astronomy and JINA: Joint Institute for Nuclear Astrophysics, Michigan
State University,\\ E. Lansing, MI 48824, USA}
\email{lee@pa.msu.edu; kenne257@msu.edu}


\begin{abstract}

Carbon-enhanced metal-poor (CEMP) stars in the halo components of the Milky Way
are explored, based on accurate determinations of the carbon-to-iron ([C/Fe])
abundance ratios and kinematic quantities for over 30000 calibration stars from
the Sloan Digital Sky Survey (SDSS). Using our present criterion that
low-metallicity stars exhibiting [C/Fe] ratios (``carbonicity'') in excess of
[C/Fe]$ = +0.7$ are considered CEMP stars, the global frequency of CEMP stars in
the halo system for \feh\ $< -1.5$ is 8\%; for \feh\ $< -2.0$ it is 12\%; for
\feh\ $<-2.5$ it is 20\%. We also confirm a significant increase in the level of
carbon enrichment with declining metallicity, growing from
$\langle$[C/Fe]$\rangle$ $\sim +1.0$ at \feh\ $= -1.5$ to
$\langle$[C/Fe]$\rangle$ $\sim +1.7$ at \feh\ $= -2.7$. The nature of the
carbonicity distribution function (CarDF) changes dramatically with increasing
distance above the Galactic plane, $|$Z$|$. For $|$Z$|$ $< 5$ kpc, relatively
few CEMP stars are identified. For distances $|$Z$|$ $> 5$ kpc, the CarDF
exhibits a strong tail towards high values, up to [C/Fe] $>$ +3.0. We also find
a clear increase in the CEMP frequency with $|$Z$|$. For stars with $-2.0 <$
[Fe/H] $< -$1.5, the frequency grows from 5\% at $|$Z$|$ $\sim 2$ kpc to 10\% at
$|$Z$|$ $\sim 10$ kpc. For stars with [Fe/H] $< -$2.0, the frequency grows from
8\% at $|$Z$|$ $\sim 2$ kpc to 25\% at $|$Z$|$ $\sim 10$ kpc. For stars with
$-2.0 <$ [Fe/H] $< -$1.5, the mean carbonicity is $\langle$[C/Fe]$\rangle$ $\sim
+1.0$ for 0 kpc $<$ $|$Z$|$ $<$ 10 kpc, with little dependence on $|$Z$|$; for
[Fe/H] $< -$2.0, $\langle$[C/Fe]$\rangle$ $\sim +1.5$, again roughly independent
of $|$Z$|$. Based on a statistical separation of the halo components in velocity
space, we find evidence for a significant contrast in the frequency of CEMP
stars between the inner- and outer-halo components -- the outer halo possesses
roughly twice the fraction of CEMP stars as the inner halo. The carbonicity
distribution also differs between the inner-halo and outer-halo components --
the inner halo has a greater portion of stars with modest carbon enhancement
([C/Fe] $\sim +0.5]$); the outer halo has a greater portion of stars with large
enhancements ([C/Fe] $\sim +2.0$), although considerable overlap still exists.
We interpret these results as due to the possible presence of additional
astrophysical sources of carbon production associated with outer-halo stars,
beyond the asymptotic giant-branch source that may dominate for inner-halo
stars, with implications for the progenitors of these populations.

\end{abstract}


\section{Introduction}

The Milky Way provides astronomers with a unique opportunity to explore the
formation and evolution of large spiral galaxies, as well as the nature of their
stellar populations and recognized structures. The key to this understanding
comes from the availability, for large numbers of individual stars, of the
powerful combination of six-dimensional phase-space information (location and
velocity) and chemical abundances. Metal-poor stars, in particular, shed light
on the early stages of galaxy formation and chemical evolution, as they
represent the fossil record of the first generations of stars that formed
shortly after the Big Bang.

Although theory suggests that the bulge of the Galaxy may harbor numerous
ancient (though not necessarily the most metal-poor) stars (e.g., Tumlinson
2010), the vast majority of presently recognized metal-poor stars are found in
the halo system of the Galaxy. According to Carollo et al. (2007; C07) and
Carollo et al. (2010; C10), the inner and outer halos possess different peak
metallicities ([Fe/H]$_{inner}$ $\sim -1.6$; [Fe/H]$_{outer}$ $\sim -2.2$), as
well as different spatial distributions, with the inner halo exhibiting a
flatter density profile than the nearly spherical outer halo. Their
analysis of the kinematics of a local sample of calibration stars from the
Sloan Digital Sky Survey (SDSS: York et al. 2000; Gunn et al. 2006)
indicated that the transition from dominance by the inner halo to the outer
halo occurs in the range 15-20 kpc from the Sun. A similar transition range
has been inferred from analysis of the ``vertical'' photometric stripes (de
Jong et al. 2010), obtained during the Sloan Extension for Galactic
Understanding and Exploration (SEGUE) sub-survey of SDSS-II (Yanny et al.
2009).

The papers by C07 and C10 also demonstrated that the inner-halo population
is essentially non-rotating, with V$_{\phi}$ = 7 $\pm$ 4 km~s$^{-1}$, while
the outer-halo population exhibits a significant retrograde signature, with
V$_{\phi}$ = $-$80 $\pm$ 13 km~s$^{-1}$ (where V$_{\phi}$ is the
Galactocentric rotational velocity). The velocity ellipsoids of these
populations differ as well, such that ($\sigma_{V_{R}}$,
$\sigma_{V_{\phi}}$, $\sigma_{V_{Z}})$ = (150 $\pm$ 2, 95 $\pm$ 2, 85 $\pm$
1) km~s$^{-1}$ for the inner halo and (159 $\pm$ 4, 165 $\pm$ 9, 116 $\pm$
3) km~s$^{-1}$ for the outer halo, evaluated in a Galactocentric
cylindrical reference frame.

The observed differences in the nature of the spatial distributions and
kinematics of the stellar populations associated with the inner- and outer-halo
components suggests that, in the context of modern hiearchical cosmogonies,
their progenitor mini-halos and subsequent merging and accretion scenarios
differed as well. If the inner halo formed from a limited number of moderately
massive mini-halos (see, e.g., Bullock \& Johnston 2005; Schlaufman et al. 2009,
2011), while the outer halo resulted from the accretion of more numerous, but
less massive ones (C07; Frebel et al. 2010; Norris et al. 2010a,b,c), one might
expect to find chemical signatures associated with the presently observed
stellar populations that reflect these differences. Previous studies have
provided hints that this may indeed be the case. For example, studies of the
[Mg/Fe] abundance ratios of stars thought to be associated with the inner halo
appear different (generally ~0.1 dex higher) than those associated with the
outer halo (Roederer 2009). This same study also demonstrated that stars
associated with the inner halo exhibit considerably lower star-to-star abundance
scatter for both the iron-peak element ratio [Ni/Fe] and the neutron-capture
element ratio [Ba/Fe] than found for stars of the outer halo. The recent study
by Nissen \& Schuster (2010) demonstrated that nearby dwarfs with halo
kinematics could be separated into two groups based on \afe. They proposed that
the high-\alp\ stars may have been born in the disk or bulge of the Milky Way
and heated to halo kinematics by merging satellite galaxies, or else were simply
members of the early generations of halo stars born during the collapse of a
proto-Galactic gas cloud, while the low-\alp\ stars may have been accreted from
dwarf galaxies. Schlaufman et al. (2009) report detections of numerous elements
of cold halo sububstructure (ECHOS) in the inner halo, essentially
overdensities in radial-velocity space along the SEGUE sightlines. The ECHOS are
systematically more Fe-rich, but less $\alpha$-enhanced than the
kinematically smooth component of the inner halo. The ECHOS are also
chemically distinct from other Milky Way components; they are more Fe-poor
than typical thick-disk stars, and both more Fe-poor and $\alpha$-enhanced
than typical thin-disk stars. See Schlaufman et al. (2011) for a more
detailed discussion.

Chemically peculiar stars, such as the $\alpha$-element-enhanced very metal-poor
star BS~16934-002 (Aoki et al. 2007a; [Fe/H] $= -2.7$), the low [Mg/Fe]
($-0.1$), high [Ca/Fe] ($+$1.1) extremely metal-poor star SDSS~J2347+0108 (Lai
et al. 2009; [Fe/H] $= -3.2$), and the low [Si/Fe] ($-1.0$), low [Ca/Fe]
($-0.6$) extremely metal-poor star HE 1424-0241, identified by Cohen et al.
(2007; [Fe/H] $\sim -4.0$), all have inferred distances (and metallicities) that
suggest membership in the outer-halo population. All three of the previously recognized
ultra metal-poor stars ([Fe/H] $\leq -4.0$, HE~0557-4840; Norris et al. 2007) or
hyper metal-poor stars ([Fe/H] $\leq -5.0$, HE~0107-5240; Christlieb et al.
2002, and HE~1327-2326; Frebel et al. 2005) are similarly thought to be members
of the outer-halo population.  Likewise, the newly discovered hyper metal-poor star
SDSS~J102915+172927 appears to have an orbit consistent with outer-halo
membership (Caffau et al. 2011). These results may all be related to the star
formation histories in the progenitor populations, their accretion histories, or
both.

In the present paper we focus on another possibly useful indicator of chemical
differences between the inner- and outer-halo components, the carbon-to-iron
ratio, [C/Fe], which we refer to as the ``carbonicity.'' In particular, we make
use of the SDSS/SEGUE calibration-star sample from SDSS DR7 (Abazajian et al.
2009) in order to search for possible contrasts between the frequency and degree
of carbon enhancement for the carbon-enhanced metal-poor (CEMP) stars from this
sample that can be kinematically associated with these two halo components.

The CEMP stars were originally defined as the subset of very metal-poor stars
([Fe/H] $\leq -$2) that exhibit elevated carbon relative to iron, [C/Fe] $>$
+1.0 (Beers \& Christlieb, 2005)\footnote{Other authors have used slightly
different criteria, e.g., [C/Fe] $> +0.7$ (Aoki et al. 2007b).}. In the last two
decades it has been recognized, primarily from spectroscopic follow-up of
metal-poor candidates selected from objective-prism surveys (e.g., Beers et al.
1985, 1992; Christlieb 2003), that roughly 20\% of stars with [Fe/H] $< -$2.0
exhibit enhanced carbonicity, up to several orders of magnitude larger than the
solar ratio (Marsteller et al. 2005; Lucatello et al. 2006). Some recent studies
(e.g., Cohen et al. 2005; Frebel et al. 2006), have claimed that this fraction
is a little lower (14\% and 9\%, respectively, for [Fe/H] $< -2.0$). The Frebel
et al. (2006) study is of particular interest, as the authors argued that the
relative fraction of CEMP stars appears to increase substantially with distance
above the Galactic plane, suggesting a possible connection with changes in the
underlying stellar populations. In any case, the fraction of CEMP stars rises to
30\% for [Fe/H] $< -3.0$, 40\% for [Fe/H] $< -3.5$, and 75\% for [Fe/H] $< -4.0$
(Beers \& Christlieb 2005; Frebel et al. 2005; Norris et al. 2007; Caffau et al.
2011 -- the new [Fe/H] $= -5.0$ star does not exhibit carbon enhancement);
definitive explanations for the origin of this increase have yet to be offered.
Regardless of the ultimate reason, these results indicate that significant
amounts of carbon were produced in the early stages of chemical evolution in the
universe.

There exist a number of classes of CEMP stars, some of which have been
associated with proposed progenitor objects. The CEMP-s stars (those with
$s-$process-element enhancement), for example, are the most commonly observed type to date.
High-resolution spectroscopic studies have revealed that around 80\% of CEMP
stars exhibit $s-$process-element enhancement (Aoki et al. 2007b). The favored
mechanism invoked to account for these stars is mass transfer of carbon-enhanced
material from the envelope of an asymptotic giant-branch (AGB) star to its
binary companion; it is this surviving companion that is now observed as a
CEMP-s star (e.g., Herwig 2005; Sneden et al. 2008; Bisterzo et al. 2011).

The class of CEMP-no stars (which exhibit no strong neutron-capture-element
enhancements) is particularly prevalent among the lowest metallicity stars (Fe/H
$< -$2.5; Beers \& Christlieb 2005; Aoki et al. 2007b). Possible progenitors for
this class include massive, rapidly rotating, mega metal-poor ([Fe/H] $< -6.0$)
stars, which models suggest have greatly enhanced abundances of CNO due to
distinctive internal burning and mixing episodes, followed by strong mass loss
(Hirschi et al. 2006; Meynet et al. 2006, 2010a,b). Another suggested mechanism
for the production of the material incorporated into CEMP-no stars is pollution
of the interstellar medium by so-called faint supernovae associated with the
first generations of stars, which experience extensive mixing and fallback
during their explosions (Umeda \& Nomoto 2003, 2005; Tominaga et al. 2007). This
model well reproduces the observed abundance pattern of the CEMP-no star
BD+44$^{\circ}$493, the ninth-magnitude [Fe/H] $= -3.7$ star (with [C/Fe] =
+1.3, [N/Fe] = +0.3, [O/Fe] = +1.6) discussed by Ito et al. (2009). The recently
reported high redshift ($z = 2.3$), extremely metal-poor Damped Lyman-$\alpha$
(DLA) system by Cooke et al. (2011; [Fe/H] $\sim -3.0$) exhibits enhanced
carbonicity ([C/Fe] $= +1.5$) and other elemental abundance signatures that
Kobayashi et al. (2011) also associate with production by faint supernovae.
It is also of interest that Matsuoka et al. (2011) have reported evidence for
strong carbon production in the early universe, based on their analysis of the
optical spectrum of the most distant known radio galaxy, TN J0924-2201, with $z
= 5.19$.

Below we seek to test if the increasing frequency of CEMP stars with declining
metallicity, and the suggested increase of the fraction of CEMP stars with
increasing distance from the Galactic plane, can be explained in the context of
an inner/outer halo dichotomy and the dominance of {\it different
carbon-production mechanisms} (the processes associated with the progenitors of the
CEMP-s and CEMP-no stars) being linked to these two populations.

This paper is outlined as follows. Section 2 describes the techniques used to
estimate the atmospheric parameters (\teff , \logg , and \feh) and carbonicity
([C/Fe]) from the low-resolution SDSS spectra, compares our estimates with a
sample of very low-metallicity stars with available high-resolution
spectroscopic determinations, and obtains first-pass estimates of the fractions
of CEMP stars for various cuts on [Fe/H]. Section 3 summarizes the
calibration-star sample from SDSS DR7 we examine here, presents the
``as-observed'' distributions of metallicity and carbonicity for this sample as
functions of height above the Galactic plane, and carries out a comparison of
the distance and rotational velocity distributions for CEMP and non-CEMP stars.
Section 4 explores the global fraction of CEMP stars of this sample in the
low-metallicity regime, describes our adopted technique for derivation of
stellar population membership probabilities for the SDSS/SEGUE DR7 calibration
stars, and obtains estimates of the fractions of CEMP stars associated with the
inner- and outer-halo populations. Finally, Section 5 summarizes our main
results and considers their implications for the formation and evolution of the
Galactic halo populations.

\section{Atmospheric Parameter Estimates and [C/Fe] Ratios \\
for the SDSS/SEGUE DR7 Calibration-Star Sample}

\subsection{Atmospheric Parameter Estimates}

Estimates of the atmospheric parameters for our program stars were obtained from
the SEGUE Stellar Parameter Pipeline (SSPP papers I-V: Lee et al 2008a,b; Allende
Prieto et al. 2008; Smolinski et al. 2011; Lee et al. 2011). Typical internal
errors for stars in the temperature range that applies to the majority of the
calibration-star sample are $\sigma (T_{\rm eff}) \sim$ 125 K, $\sigma ({\log
g}) \sim $ 0.25 dex, and $\sigma {\rm [Fe/H]} \sim$ 0.20 dex. The external
errors in these determinations are of similar size, as discussed in the SSPP
references listed above.

In C10, a correction was applied for the metallicity determinations of the SSPP,
which we adopt here as well:

\begin{equation}
{\rm [Fe/H]}_C = -0.186 + 0.765*{\rm [Fe/H]}_A - 0.068*{\rm [Fe/H]}_A^2
\end{equation}

\noindent where [Fe/H]$_A$ is the adopted metallicity from the SSPP, and
[Fe/H]$_C$ is the corrected metallicity. This polynomial has little effect on
stars with metallicity greater than about [Fe/H] = $-2.5$, but lowers the
estimated metallicities for stars below this abundance by 0.1 to 0.2 dex, an
offset that was shown to exist between the DR7 SSPP-derived metallicities and
previous high-resolution spectroscopic measurements.

\subsection{Estimation of Carbon Abundance Ratios}

Carbon-to-iron abundance ratios ([C/Fe]) are estimated from the CH G-band
at $\sim$ 4300~{\AA} by matching the observed SDSS spectra near this
feature with an extensive grid of synthetic spectra. In order to construct
the grid we employed the NEWODF models of of Castelli \& Kurucz (2003).
Synthetic spectra were generated using the \texttt{turbospectrum} synthesis
code (Alvarez \& Plez 1998), which employs line broadening according to the
prescription of Barklem \& O'Mara (1998) and Barklem \& Aspelund-Johansson
(2005). The molecular species CH and CN are provided by B. Plez (private
communication; Plez \& Cohen 2005). The other linelists used are the same
as in Sivarani et al. (2006). For the purpose of this exercise we adopted
the solar abundances of Asplund et al. (2005).

The synthetic spectra cover wavelengths between 3600{\AA} and 4600{\AA}, with
original resolution of $\Delta$$\lambda$ = 0.005{\AA}, smoothed to the SDSS
resolving power ($R = 2000$) and re-binned to linear 1{\AA} pixels. The stellar
parameters of the grid cover the ranges 3500~K $\le $ \teff\ $\le 9750$~K (steps
of $250$~K), $0.0 \le$ \logg\ $\le 5.0$ (steps of 0.5 dex) and $-2.5 \le$ [Fe/H]
$\le $ 0.0 (steps of $0.5$ dex). For stars with [Fe/H] $< -2.5$, models with
[Fe/H] $= -2.5$ were adopted. At the time this analysis was carried
out, lower metallicity models from this grid were not available. However, we did
have sparsely-spaced carbon-enhanced models generated by B.Plez, which extended
down to [Fe/H] $= -5.0$, for 4000~K $\le$ \teff\ $\le$ 6000~K. We carried out a
number of tests of our use of the [Fe/H] $= -2.5$ models extrapolated to lower
metallicities, which produced results in good agreement with Plez's models. Our
program stars include objects with temperatures above 6000~K, where any effects
due to extrapolation to lower metallicities will be lower still. Indeed Masseron
et al. (2005) points out that models with \teff\ $>$ 6000~K are not affected by
enhanced carbon. In any event, our past experience using models of lower
metallicity from other sources has indicated that very little changes when
dropping below [Fe/H] $= -2.5$. This is also indicated by the generally
excellent agreement of our [C/Fe] determinations with the high-resolution
results discussed below. The carbon abundance in the grid goes from [C/H] =
[Fe/H] $-$ 0.5 to [C/H] = +0.5 (the upper limit is in agreement with AGB models)
. For example, at [Fe/H] = $-$2.5, the grid covers the range from [C/H] = $-$3.0
to $+$0.5, which corresponds to the range of carbonicity $-0.5 <$ [C/Fe] $<$
$+$3.0.

Once constructed, we linearly interpolate within this grid, which is sufficient
for the size of the steps in the parameters. We have checked this by taking a
worst-case scenario, generating test synthetic spectra at low temperatures
(\teff\ $= 3500$~K) and over the above ranges of gravity, metallicity, and
carbon abundance. The linearly interpolated grid was able to recover the input
parameters to within a few tenths of dex, which is consistent with our expected
errors in the method. Estimation of carbon abundance was accomplished using
chi-square minimization of the deviations between the observed and synthetic
spectra in the wavelength region between 4285 {\AA} and 4320 \AA, as carried out
by the IDL routine AMOEBA (a down-hill Simplex search procedure). The initial
value for [C/H] for the the global grid search was set to the same value as the
input stellar [Fe/H] ([C/Fe] = 0.0). During the search, the carbon abundances
are allowed to vary; all other stellar parameters are kept constant. In order to
provide some protection from falling into local minima, separate searches were
performed with lower and higher ranges of [C/Fe] considered. In almost all cases
these converged to the same minima as found for the global search. When not, we
took the value that resulted in the best match in the region of the CH G-band,
as judged from a correlation coefficient for the resulting match.

Figure 1 provides an example of the spectral matching process for determination
of [C/Fe] for a warm CEMP star in our sample. The upper panel shows the input
optical spectrum (black line) superposed with a synthetic spectrum with [C/Fe] =
0 (red line). The middle panel shows the best matches to the CH G-band obtained
from the three different ranges of [C/Fe] considered. The lower panel shows the
final adopted match.

Note that our procedures are not traditional synthesis analyses, but are based
on spectral matching. As part of this approach, the continuum-flattened observed
spectra must be registered to match similarly flattened synthetic spectra.
Because rectification of the stellar (and synthetic) continua is sometimes
imperfect, small deviations over localized regions of spectral range can
occasionally appear. Careful inspection of the bottom panel of Figure 1 reveals,
for example, a slight mismatch near the red end. We have taken care to minimize
the occurrence of these mismatches to the extent feasible, with particular
effort made in the range that is explored for performing the match to the CH
G-band. In any case, since the atmospheric parameters are set before conducting
the matching, slight registration offsets outside of the CH G-band region have
no effect on our derived [C/Fe]. We note that most of the weak features in the
region of the CH G-band are real, and not due to noise.

Similar procedures have been employed in previously published
papers (Beers et al. 2007; Kennedy et al. 2011), to which we refer the reader for
additional discussion of these techniques. We have also pursued a modest sanity
check by carrying out full syntheses for a small number of spectra using the
approach employed by Norris et al. (2010c), based on the synthesis code of
Cottrell \& Norris (1978). These comparisons indicate that we are able to
replicate derived [C/Fe] by our spectral matching approach to within 0.1-0.2
dex, at the one-sigma level in the precision for the majority of our
program stars.

Estimates of the errors in the final determinations of [C/Fe] were obtained
based on a set of noise-injection experiments for stars over a range of
temperatures and S/N ratios. These experiments indicated that, for stars with a
minimum S/N $= 15/1$ in the region of the G-band, [C/Fe] could be measured over
the \teff\ range of our sample with a maximum error of 0.5 dex, decreasing to
0.05 dex for the highest S/N spectra (S/N $> 50/1$). We assigned final errors to
the estimate of [C/Fe] using a linear function in S/N between these extremes.
Spectra not achieving the minimum S/N level (or which suffered from anomalies
such as pixel dropouts) were considered non-observations (i.e., they were
dropped from the sample). In addition, in order to claim a detection we required
that the equivalent width of the CH G-band obtain a minimum value of 1.2 \AA, a
value again chosen based on inspection of the noise-injection
experiments.\footnote{The CH G-band equivalent width measured in this study
covers a wider wavelength region than that defined by Beers et al.
(1999): 26 {\AA} (this paper) vs. 15 {\AA} (Beers et al. 1999), but
centered on the same wavelength (4305 \AA). It also makes use of a global fit to
the continuum, rather than the fixed sidebands employed previously. Thus,
although the equivalent widths are similar, they are not identical.} In cases
where this condition was not met, we consider the measured [C/Fe] an upper
limit.

\subsection{Comparison with High-Resolution Spectroscopic Estimates}

We obtain a check on our determinations of atmospheric parameters and
[C/Fe], for at least a subset of our stars, based on available high-resolution
follow-up spectra. There are three sources for our comparisons:  Aoki et al.
(2008), Behara et al. (2010), and Aoki et al. (2011, in prep.). Since these
high-resolution spectroscopic programs were mostly interested in very and
extremely metal-poor stars, our comparison sample is dominated by stars with
[Fe/H] $< -2.5$. Table 1 lists the derived atmospheric parameters and carbon
abundance ratios for the stars in common (the label HIGH indicates the
high-resolution results, while the label SSPP refers to our analysis of the
low-resolution SDSS spectra). There are 23 unique objects listed, although one
star has only upper limits for [C/Fe] determined. Note that four stars in this
table were reported on by two sets of authors, which provides some feeling for
the level of systematic differences in the derived parameters. The Aoki et al.
(2011) paper, which supplies information for most of our comparison
sample, concluded that the temperature estimates provided by the SSPP were
sufficiently good that they simply adopted them for their analysis (although
they tested alternative methods, they could not improve upon the SSPP results).
As a result, a comparison of \teff\ determinations for the high-resolution and
low-resolution spectroscopic analyses cannot be fairly carried out. In the case
of stars that were observed by multiple groups, straight averages of the listed
parameters are used (except when the Aoki et al. 2011 {\it assigned} value of \logg\ $=
+4.0$, as described below, can be replaced by a {\it measured} value from the other
sources).

Figure 2 shows a comparison of the derived parameter estimates. Robust
estimates of the zero-point offsets and rms scatter indicate good agreement
in metallicity ($\langle \Delta {\rm [Fe/H]}\rangle = -0.09$~dex;
$\sigma({\rm [Fe/H]}) = 0.27$~dex), close to the random errors expected for
carbon-normal stars from the SSPP (0.2 dex). Surface gravity exhibits an
acceptably small zero-point offset but a larger scatter ($\langle \Delta
(\log g)\rangle = -0.27$~dex; $\sigma(\log g) = 0.66$~dex). Larger
differences in the \logg\ determinations might be expected for several
reasons. First, the comparison spectra for the majority of our program
stars (18 of 23) are based, at least in part, on ``snapshot''
high-resolution spectra (i.e., lower S/N) spectra reported by Aoki et al.
(2011). For these spectra, estimates of \logg\ were fixed to \logg\ = 4.0
for all stars with \teff\ $>$ 5500 K. This was done because the usual
procedure of matching Fe abundances based on Fe I and Fe II lines was not
uniformly possible for the warmer stars, due to the weakness of the Fe II
lines and the less than ideal S/N. According to Aoki et al. (2011), the
true surface gravities for such stars could lie anywhere in the range
\logg\ = 3.5 to \logg\ = 4.5.  Note that the reported surface-gravity
offset and scatter above does not include the three Aoki et al. (2011)
stars that did not have reported estimates from other sources.

Surface gravity estimates from the high-resolution spectra of giants with \teff
$ < 5500$~K are determined by Aoki et al. (2011) based on analysis of the Fe I
and Fe II lines, in the usual manner. Note that Aoki et al. also identified four
stars in this temperature range to be cool main-sequence stars, rather than
giants, based on the weak lines of their ionized species. For these stars,
gravity estimates are obtained by matching to isochrones for metal-poor
main-sequence stars (a similar process to that carried out by Aoki et al. 2010).

It is expected that the high-resolution analysis estimates of \logg\ for the cooler
stars are precise at a level no better than 0.3 dex. Secondly, difficulty in
estimation of \logg\ for CEMP stars might not be suprising, given that molecular
carbon bands (in particular for later type stars) can easily confound
gravity-sensitive features in a low-resolution spectrum. However, in our
analysis we have taken care to avoid regions of the spectrum where molecular
carbon bands have corrupted the gravity-sensitive features. In reality, this
problem is only encountered for the coolest stars with the strongest molecular
bands, which are a distinct minority of our sample. Our sample only includes
stars in the temperature range 4500~K $<$ \teff\ $<$ 7000 ~K, and among the 813
carbon-rich stars considered in our analysis below, none of them have \teff\ $<$
5000~K, and only 10\% have \teff\ $<$ 5750~K. Taking a global approximation that
the high-resolution determinations of surface gravity contribute 0.4 dex to the
rms scatter comparison with the low-resolution estimates, this indicates that
the SSPP estimates for these stars have an external error of determination of
$\sqrt{(0.66^2 - 0.4^2)}$ = 0.52 dex. We also note that the great majority of
the stars analysed in our sample have [Fe/H] $>$ $-$2.5, for which surface
gravity estimates should be better determined, due to the increasing strength of
their gravity-sensitive metallic features.

The agreement in estimates of the carbonicity is quite good ($\langle \Delta
{\rm [C/Fe]}\rangle = +0.05$~dex; $\sigma({\rm [C/Fe]}) = 0.29$~dex), since we
expect even the high-resolution determinations to be precise to no better than
about 0.15-0.20 dex. This suggests that our external errors for estimates of
[C/Fe] are on the order of 0.25 dex. Note that this level of agreement between
[C/Fe] determinations based on the high-resolution and SSPP analyses would not
be possible if the rough estimates of \logg\ had a strong influence on our
procedures. However, at the suggestion of an anonymous referee, we have carried
out explicit tests of the effect of incorrect surface-gravity determinations on
the derived [C/Fe].

In order to test the impact on the derived [C/Fe] from the adoption of an incorrect
surface gravity, we have used synthetic spectra (with known atmospheric
parameters and [C/Fe]) from our grid covering three fixed metallicities, [Fe/H]
= $-1.0$, $-1.5$, and $-2.5$, three fixed gravities, \logg\ = 2.0, 3.0, and 4.0,
which spans the range of the majority of our program stars, and two levels of
carbonicity, [C/Fe] = 0.0 and [C/Fe] = +1.5. We then intentionally perturbed their
input \logg\ values by $-1.0$ dex, $-0.5$ dex, $+0.5$ dex, and $+1.0$ dex, and
derived estimates of their [C/Fe] following the procedures described above. As
can be seen from inspection of Figure 3 (which shows the case for [C/Fe] = 0.0),
the effects on estimates of [C/Fe] never exceed 0.5 dex (and then only in the
most extreme cases of $\pm$ 1.0 dex variation in \logg\ ), and are generally much
smaller than that. Not surprisingly, the largest variations occur for the
warmest stars, which have weaker CH G-band features. However, our sample
included only a small fraction of stars with \teff\ $>$ 6500~K ($\sim 3$ \%), so
this is not expected to have a major effect. The mean zero-point offsets and rms
variations in the derived [C/Fe], relative to the known value across all \teff\
and \logg\ considered, are shown for each panel in Figure 3, and are acceptably
small. Similar results apply to the case when we fix the input [C/Fe] = +1.5.
Figure 3 considered the solar [C/Fe] case, since we are more concerned with
false positives that would cause us to count a star as carbon-enhanced when it
is not. We conclude that, while some sensitivity to estimates of [C/Fe] may
exist due to errors in estimates of \logg\ , its impact on our results should be
minimal, except for truly extreme cases.

\subsection {Detections, Upper Limits, and First-Pass Frequencies of CEMP Stars}

In this work there are 31187 unique stars for which estimates of [C/Fe]
were carried out (other objects were either repeats, in which case they
were averaged, had insufficient spectral S/N ratios, were clear cases of
QSOs, cool white dwarfs, or very late-type stars, or had some spectral
defect in the region of the CH G-band which prevented measurements being
obtained -- all such stars are dropped from the subsequent analysis). The
remaining sample can be divided into two categories: stars that have a
measured [C/Fe] (with the G-band detected; N$^{D}$ = 25647), and stars for
which only an upper limit on [C/Fe] has been obtained (G-band undetected;
N$^{L}$ = 5540). We call the associated two sets of stars subsample D and
subsample L, respectively. Figure 4 shows [C/Fe] as a function of
metallicity for these two categories. The ``ridge lines" in both of the
panels are due to grid effects in the chi-square matching procedure. It is
worth noting that most of the stars with measured [C/Fe] (subsample D)
exhibit carbonicity below [C/Fe] = +0.7, which indeed appears to be a
natural dividing threshold between carbon-normal stars and carbon-rich
stars. This limit was also established by Aoki et al. (2007b), who also
included in their analysis expected evolutionary effects on the definition
of CEMP stars. We adopt this value of [C/Fe] to define the carbon-rich stars in
our sample, without making any luminosity correction adjustment, since our
sample includes very few cool giants. There is an evident correlation
between [C/Fe] and [Fe/H] seen in this figure, in particular for stars with
[Fe/H] $< -$ 1.0. The relative number of stars with carbon excesses
increases as the metallicity decreases, as does the level of carbonicity,
as reported previously by several studies based on much smaller samples of
stars.

Adopting the definition of carbon-rich stars as above, we distinguish two
subcategories within subsamples D and L: stars with [C/Fe] $\leq$ +0.7 (C-norm),
and stars with [C/Fe] $>$ +0.7 (C-rich). In the case of subsample L, the carbon
status for a star can be assessed with certainty only when [C/Fe] $\leq$ +0.7,
and remains unknown for stars having [C/Fe] $>$ +0.7. Indeed, suppose that the
limit assigned to a given star is [C/Fe]$_{lim}$ = +1.5. Then, all values below
[C/Fe]$_{lim}$ can still be accepted for that star, including carbonicity
well below [C/Fe] = +0.7. When the upper limit is below [C/Fe] = +0.7, however,
the carbon status of the star can be assessed as C-norm. Not surprisingly, there
is a strong temperature effect in the assessment of the carbon status for stars
in our sample. For example, stars with higher \teff\ (above \teff\ $\sim
6250$~K) would require quite high [C/Fe] for the CH G-band to be detected; stars
without a detected CH G-band are included in the L subsample.  Thus, the fractions
of CEMP stars we derive in this paper are lower limits to the true fractions.
In a future paper we plan to obtain an explicit correction function to account
for the ``missing'' stars due to this temperature effect, in order to assess its
impact on the derived frequencies of CEMP stars.

Taking into account the above definitions, the fraction of C-rich stars can
be formulated as:

\begin{equation}
F_{C-rich} = \frac{N^{D}_{C-rich}}{N^{D}_{C-rich} + N^{D}_{C-norm} + N^{L}_{C-norm}}
\end{equation}

\medskip

\noindent where $N^{D}_{C-norm}$ and $N^{D}_{C-rich}$ are stars belonging to subsample D
and having [C/Fe] $\leq$ +0.7 and [C/Fe] $>$ +0.7, respectively, while
N$^{L}_{C-norm}$ are stars in subsample L and having [C/Fe] $\leq$ +0.7. Stars with
unknown carbon status are not included in the above definition.

Table 2 reports the number of stars belonging to the various categories for
different ranges of metallicity, as well as the total fractions of C-rich
stars.  Note that, when comparing the reported fractions in this table with other
fractions reported in this paper, the Table 2 fractions have no restriction
on whether or not acceptable proper motion and radial velocity measurements were
available for a given star.

\section{The Nature of the Carbon-Rich Star Sample}

\subsection{Radial Velocities, Distance Estimates, and
Definition\\ of the Extended and Local Samples}

We begin with the 31187 unique DR7 calibration stars for which estimates of
[C/Fe] exist, and with distances slightly revised from those presented in C07
and C10, as described below. Details concerning the nature of the
calibration-star sample can be found in C10; here we recall a few facts
concerning these stars that are of importance for the present analysis.

Spectra of the SDSS/SEGUE calibration stars were obtained to perform
spectrophotometric corrections and to calibrate and remove the night-sky
emission and absorption features (telluric absorption) from SDSS spectra. The
spectrophotometric calibration stars cover the apparent magnitude range 15.5 $<
g_{0} <$ 17.0, and satisfy the color ranges 0.6 $< (u-g) _{0} <$ 1.2 ; 0.0 $<
(g-r) _{0} <$ 0.6.\footnote{The subscript 0 in the magnitudes and colors
indicates that they are corrected for the effects of interstellar absorption and
reddening, based on the dust maps of Schlegel et al. (1998).} The telluric
calibration stars cover the same color ranges as the spectrophotometric
calibration stars, but at fainter apparent magnitudes, in the range 17.0 $<
g_{0} <$ 18.5.

The C10 paper describes the radial velocity estimates (which have a precision of 5-20
km~s$^{-1}$, depending on the S/N ratio of the spectrum, and with a negligible
zero-point offset), as well as photometric distance estimates, obtained by using
the SSPP surface-gravity estimate for luminosity classification, then following
the procedures of Beers et al. (2000). Since the estimated \logg\ is used only
for classification, this means that for stars redder than than main-sequence
turnoff (MSTO), the expected large differences in surface gravity for dwarfs and
giants make even approximate \logg\ estimates sufficient -- errors of 1 to 2 dex
in \logg\ would have to be routinely made in order to confound this procedure.
Close to the MSTO, any method of photometric distance estimation becomes more
problematic, but the difference in the derived distances for stars just above or
just below the MSTO decreases the closer they are to it, which mitigates against
these effects.

Sch\"onrich et al. (2011) have criticized the Beers et al. (2000) method of
distance determination, and the results of the C07 and C10 that relied on it.
They claimed that the counter-rotating halo found in C07 and C10 is a result of
biases in distance estimates for main-sequence dwarfs, and furthermore that the
distance derivation is influenced by sorting a subset of the stars into
incorrect positions in the color-magnitude diagram of an old, metal-poor
population. In a rebuttal paper, Beers et al. (2011) demonstrated that the
the Sch\"onrich et al. claims concerning dwarf distances are incorrect (due to
their adoption of the wrong main-sequence absolute magnitude relationship from
Ivezi\'c et al. 2008). Furthermore, the claimed retrograde tail in the
rotation-velocity distribution was shown to arise from the measured
asymmetric proper motions, and was not the result of the propagation of
distance errors. In any event, we have applied the procedure suggested in
the rebuttal paper to reassign intermediate-gravity stars that were
originally classified as main-sequence turnoff stars into dwarf or
subgiant/giant luminosity classifications, in cases where their derived
\teff\ were substantially lower than expected for the turnoff. Revised
distances for these stars have been adopted as well. A reassessment of the
kinematics indicates that the retrograde signal for the subsample at very
low metallicity ([Fe/H] $< -$2.0) remains. The interested reader can find
more details in Beers et al. (2011). Based on the discussion in that paper,
we believe that our distances should be accurate to on the order of
15-20\%.

The C10 procedures applied a series of cuts to their sample designed to better
enable measurement of the kinematic and orbital properties of the various
stellar populations considered. This produced a subsample of stars in the local
volume (distance from the Sun $d < 4$ kpc, and with 7 kpc $< R <$ 10 kpc, where
$R$ is the Galactocentric distance projected onto the plane of the Galaxy),
which they referred to as the ``Local Sample" (N $\sim$ 17000). These
restrictions were made in order to mitigate against the increase in the errors
in the derived transverse velocities, which scale with distance from the Sun,
and to improve the applicability of the simple models for the adopted form of
the Galactic potential. For our present analysis we do not need to redetermine
the velocity parameters of the underlying stellar populations, so we can relax
these cuts somewhat. As described below, in order to increase the numbers of
stars in our sample with measured [C/Fe], we have changed the constraint on the
distance from the Sun, from $d < $ 4 kpc to $d <$ 10 kpc, and removed the
constraint on the projected distance $R$. With these relaxed cuts the total
number of stars climbs to 30874. We refer to this new sample as the ``Extended
Sample."

The upper panel of Figure 5 shows the as-observed metallicity distribution
function (MDF) for the DR7 calibration stars belonging to the Extended Sample
(black histogram), as well as for stars in the C10 Local Sample, as described
above (red histogram). We use the term ``as-observed'' in order to call
attention to the fact that the selection functions for the calibration stars
were not intended to return a fair sample of stars, suitable for an unbiased
analysis of the distribution of metallicity for stars in the Galaxy. Rather, the
calibration stars were selected to emphasize the numbers of moderately
low-metallicity stars that might serve to best constrain the spectrophotometry
and telluric line corrections carried out as part of the SDSS spectroscopic
pipeline reductions. Thus, these MDFs are a ``sample of convenience,''
one that is still useful for providing guidance as to the presence of various
stellar populations in metallicity space, but not for obtaining estimates of
their relative normalizations (for which other samples drawn from SDSS are more
suitable). No selection for or against carbon-enhanced stars was carried out in
the selecton of the calibration-star sample.

The MDFs of the two samples are clearly very similar, and comprise stars with
metallicities that sample all of the primary stellar components of the Galaxy
(with the exception of the bulge). The lower panel of Figure 5 is the
distribution of [C/Fe] (estimated as described above), which we refer to as the
carbonicity distribution function (CarDF) for the Extended Sample (black
histogram), and for the Local Sample (red histogram) \footnote{We employ the
term ``carbonicity distribution function'' to emphasize that we are describing
the carbon-to-iron abundance ratio, rather than the carbon abundance
distribution itself.}. The inset in the panel shows a rescaling appropriate for
the high [C/Fe] tail of this distribution. As in the case of the MDFs, the shape
of the two CarDFs are similar, with a strong peak at [C/Fe] $\sim$ +0.2 to +0.3, and two
tails, a weak one at low carbonicity ([C/Fe] $<$ 0), and a strong
one that extends to high carbonicity, +0.5 $<$ [C/Fe] $<$ +3.0. The total number
of stars with high carbonicity is significantly different in the two samples.
Indeed, for the Extended Sample we find 728 stars with [C/Fe] $>$ +0.7, while
for the Local Sample the number is reduced to 318.

\subsection{As-Observed MDF and CarDF of the Extended Sample as a Function of
Distance from the Galactic Plane}

We now examine the MDFs and the CarDFs of the Extended Sample of calibration
stars for different intervals in $|$Z$|$, with cuts chosen to ensure there
remain adequate numbers of stars in each interval.

In Figure 6, the first (left-hand) column and the third column show the MDFs,
while the second column and the fourth (right-hand) column are the CarDFs. In
the first and third columns, the red arrows indicate the peak metallicities of
the various stellar populations considered by C10. In the second and fourth
column, the blue arrows show the location of the solar carbon-to-iron ratio
([C/Fe] = 0.0), and the location of the natural threshold that divides
carbon-normal stars from carbon-rich stars ([C/Fe] = +0.7), as identified above.

Examination of the first column of panels in Figure 6 shows how the MDF changes
from the upper-left panel, in which there are obvious contributions from the
thick-disk, the metal-weak thick disk (MWTD), and inner-halo components in the
cuts close to the Galactic plane, to the lower-left panel, with an MDF dominated
primarily by inner-halo stars. In the third column of panels, with distances
from the plane greater than 5 kpc, the transition from inner-halo dominance to a
much greater contribution from outer-halo stars is clear. This demonstration is,
by design, independent of any errors that might arise from derivation of the
kinematic parameters, and provides confirmation of the difference in the
chemical properties of the inner- and outer-halo populations originally
suggested by C07. The second and fourth columns show the results of the same
exercise for the CarDFs. Close to the Galactic plane (second column; up to
$|$Z$|$ = 3 kpc), where the thick disk and MWTD are the dominant components, the
CarDF is strongly peaked at values between [C/Fe] = 0.0 and [C/Fe] = +0.3. The
CarDFs of the thick disk and MWTD will be explored in a future paper. Here we
simply note that in the regions close to the plane, where these components
dominate the sample, there are not many stars populating regions of
high carbonicity. At larger distances from the Galactic plane, where the inner
halo begins to be the dominant component, there appears a tail in the
carbonicity distribution towards higher values, [C/Fe] $>$ +0.5. In the fourth
column of panels, where the distance from the Galactic plane is $|$Z$|$ $>$ 5
kpc, and where we expect to see the beginning of the transition from inner-halo
dominance to outer-halo dominance, the tails towards high [C/Fe] values become
even more evident. The CarDF exhibits a strong peak at [C/Fe] $\sim$ +0.3 and a long
tail towards high values of carbonicity, up to [C/Fe] = +3.0. The nature of the
CarDF is likely to be influenced by the change of the MDF as a function of the
distance from the Galactic plane, due the well-known trend of increasing [C/Fe]
with declining [Fe/H]. However, as discussed below, there is evidence that the
observed changes may reflect {\it real differences} in the chemistry of the
inner- and outer-halo populations, even at a given (low) metallicity.

By adopting the threshold of [C/Fe] $> +0.7$, the fraction of SDSS/SEGUE
calibration stars with high carbonicity in the subsample at $|$Z$|$ $>$ 9
is 20\%, (which, as argued above, is a lower limit), in line with previous
estimates for stars with [Fe/H] $< -2.0$.

\subsection{Comparisons of Distance and Rotational Velocity
Distributions\\ for CEMP and non-CEMP Stars}

We now consider how the nature of our Extended Sample differs for the
low-metallicity ([Fe/H] $< -1.5$) CEMP ([C/Fe] $> +0.7$) and non-CEMP ([C/Fe] $<$
+0.7]) stars. Figure 7 shows two columns of panels, corresponding to the stars
in our sample with different cuts on Z$_{max}$ (the maximum distance of a
stellar orbit above or below the Galactic plane); the left-hand column includes
stars at all Z$_{\rm max}$, while the right-hand column only include stars
satisfying Z$_{\rm max}$ $>$ 5 kpc. The top panels show the distributions of
distance, $d$, while the bottom panels are the distribution of the
Galactocentric rotational velocity, V$_{\phi}$. For all panels, CEMP stars are
shown as red dot-dashed histograms; non-CEMP stars are shown
as black solid histograms.

As can be seen in the top panel of each column, the peak of the distance
distribution in both cases is, for the non-CEMP stars, $d \sim 2$ kpc, while for
the CEMP stars, a softer peak is seen around $d \sim$ 2.5-4 kpc (closer
to $d \sim$ 4 kpc for the higher Z$_{\rm max}$ cut), with long tails extending
to the 10 kpc cutoff of the Extended Sample. At all distances beyond $d \sim$ 3
kpc the relative fraction of CEMP stars exceeds that of the non-CEMP stars,
while closer than $d \sim$ 3 kpc the relative fraction of non-CEMP stars is
greater than that of the CEMP stars. A Kolmogorov-Smirnov (K-S) test of the
distance distributions indicates that the hypothesis that the CEMP and non-CEMP
stars are drawn from the same parent population is rejected at high significance
($p < 0.001$) for both cuts on Z$_{\rm max}$.

For the case of Z$_{max} > 0$ kpc, stars classified as dwarfs represent 72\% of
the non-CEMP sample and 42\% of the CEMP sample, while subgiants and giants
represent 24\% of the non-CEMP sample and 50\% of the CEMP sample. For both the
non-CEMP and CEMP samples, main-sequence turnoff stars represent less than 10\%
of the samples. For the case of Z$_{max} > 5$ kpc, dwarfs comprise 52\% of the
non-CEMP sample and 29\% of the CEMP sample, while subgiants/giants represent
44\% of the non-CEMP sample and 63\% of the CEMP sample. Again, main-sequence
turnoff stars comprise less the 10\% of both samples. Thus, the difference in
distance distribution is not solely dependent on the luminosity classes
associated with the populations split on carbonicity; dwarfs and
subgiants/giants are both significantly represented in each of the non-CEMP and
CEMP samples.

From inspection of the bottom panels of Figure 7, the non-CEMP stars exhibit a
slightly asymmetric distribution of rotational velocities centered close to
V$_{\phi} \sim 0$ km s$^{-1}$, and a weak retrograde tail. By contrast, the CEMP
stars exhibit a rather strong asymmetry extending to large retrograde
velocities. A K-S test of the rotational-velocity distributions indicates that
the hypothesis that the CEMP and non-CEMP stars are drawn from the same parent
population is rejected at high significance ($p < 0.001$) for both cuts on
Z$_{\rm max}$.

It is interesting to consider the distribution of luminosity classes for the
split on [C/Fe] for the retrograde stars with V$_{\phi} < -100 $ km s$^{-1}$.
For the case of Z$_{\rm max} > 0$ kpc, stars in the retrograde tail classified as
dwarfs represent 66\% of the non-CEMP sample and 20\% of the CEMP sample, while
subgiants/giants represent 30\% of the non-CEMP sample and 73\% of the CEMP sample. For
both non-CEMP and CEMP samples main-sequence turnoff stars represent less than
10\% of the samples. For the case of Z$_{\rm max} > 5$ kpc, the retrograde stars
classified as dwarfs comprise 48\% of the non-CEMP sample and 18\% of the CEMP
sample, while subgiants/giants represent 47\% of the non-CEMP sample and 75\% of
the CEMP sample. Again, main-sequence turnoff stars comprise less than 10\% of
both samples. Significant fractions of dwarfs and giants are present in the
retrograde tail for both the non-CEMP and CEMP samples. Thus, the retrograde
signature is unlikely to be due to a preponderance of stars with aberrant
distance estimates owing to luminosity misclassifications.

If we specialize to the highly retrograde tails, we find that for \zmax\  $> 0$
kpc, the portion of the non-CEMP sample in this tail is only 8\% for \vphi\ $<
-150$ \kms\ and 4\% for \vphi\ $< -200$ \kms\ . For the CEMP stars, these
portions are 17\% for \vphi\ $< -150$ \kms\ and 12\% for \vphi\ $< -200$ \kms\ ,
respectively. For the case of \zmax\  $> 5$ kpc, the portion of the non-CEMP
sample is 12\% for \vphi\ $< -150$ \kms\ and 8\% for \vphi\ $< -200$ \kms\ . For
the CEMP stars, these portions are 24\% for \vphi\ $< -150$ \kms\ and 19\% for
\vphi\ $< -200$ \kms\ , respectively. Thus, in both cases, a split on the level
of carbon enhancement leads directly to rather different relative population of
the retrograde tails, compared to the full distributions of non-CEMP and CEMP
stars. Identifying the highly retrograde tails with the outer-halo population,
we can already infer that the outer-halo component appears to harbor a greater
fraction of CEMP stars than the inner-halo component.

In addition, before considering a more complete discussion below, we can
make use of the \vphi\ distribution to obtain an estimate of the approximate
fraction of CEMP stars in the outer halo. We proceed by asserting that the stars
in the highly retrograde tail, with \vphi\ $< -200$ \kms\ , are very likely to be
members of the outer-halo population. This follows because the
dispersion in \vphi\ for an essentially non-rotating inner halo is on the order
of 100 \kms\ , and placing a cut at two sigma below the mean rotation of the
inner halo excludes all but 2.5\% of likely inner-halo stars. With this
assumption, we find that 11\% of the stars in the highly retrograde tail (and
with \zmax\  $>$ 5) kpc) are CEMP stars. This differs from the calculations
immediately above, in that we are considering the fractions of CEMP stars
relative to the total number of stars (including those from the L subsample, but
not those with unknown carbon status; i.e., we are using Eqn. 2). Recall that
this calculation applies only for stars with [Fe/H] $< -1.5$.  For this
subsample we find a mean carbonicity of $\langle$[C/Fe]$\rangle$ = +1.47 $\pm$
0.07, where the error is the standard error of the mean.  This value can be
taken as a first-pass estimate of the mean outer-halo carbonicity.

Figure 8 shows the observed distributions of the measured proper motions in
the right ascension and declination directions for the low-metallicity (\feh $<
-1.5$) stars in our sample, for two cuts on \zmax, represented by small blue
dots. The same panels show, represented as red stars, the objects that populate
the highly retrograde tails of the distribution of rotational velocity shown in
Figure 7. Note that cuts at \vphi\ $< -$150 \kms\ and \vphi\ $< -$200 \kms\ are
shown in the upper and lower rows of panels, respectively. As can be appreciated
from inspection of this figure, the proper motions in both directions for the
stars assigned to the highly retrograde tail are asymmetrically distributed, and
explore much larger values, relative to the rest of the sample. This supports
the reality of the highly retrograde signature seen in Figure 7, and indicates
that it is due primarily to the proper motions of the participating stars. See
Beers et al. (2011) for additional details concerning the veracity and
interpretation of SDSS proper motions for the SDSS calibration-star sample.

\section{Extreme Deconvolution and Membership Probabilities}

In the relatively nearby volume explored by the SDSS/SEGUE calibration-star
sample, C07 and C10 have shown that the two stellar components of the Galactic
halo are strongly overlapped in their spatial distribution, velocity ellipsoids,
and MDFs. Thus, to explore possible differences in the frequencies and mean
carbonicities of CEMP stars, it is essential to assign inner- and
outer-halo membership probabilities to each star in the sample that is likely to
belong to the halo system. We describe how this is accomplished in the sections
below.

\subsection{Basic Parameters}

In the solar neighborhood, the set of parameters that best identify the presence
of the main structures are the distance from the Galactic plane, the rotational
velocity of a star in a cylindrical frame with respect to the Galactic center,
and the stellar metallicity (see C10). For consideration of the disk system of
the Galaxy, the distance from the Galactic plane is best represented by $|$Z$|$
(the present distance of a star above or below the Galactic plane), while for
the halo components, a more suitable distance is Z$_{\rm max}$, which depends on the
adopted gravitational potential. The choice of Z$_{\rm max}$ for the Galactic halo
is necessary because of the much larger spatial extent of its two primary
components, relative to that of the disk components.

Before seeking a deconvolution of the inner- and outer-halo components, we need
to check for possible correlations between all of the basic parameters we wish to
employ. In C10 we have shown that Z$_{\rm max}$ has a significant correlation with
V$_{R}$ and V$_{Z}$, but not with V$_{\phi}$, other than that expected from the
presence of the thick-disk population at high positive rotation velocity and the
halo at lower rotational velocity. A very similar behavior is confirmed for the
Extended Sample as well (Figure 9). We conclude that the Galactocentric
rotational velocity, V$_{\phi}$, and the vertical distance, Z$_{\rm max}$, when
combined with metallicity, can be used to obtain useful information on the
different stellar populations present in the Extended Sample.

The primary chemical parameter of the present analysis is the carbonicity,
[C/Fe], so it is important to check for its possible correlations with the
basic parameters defined above. The result of this exercise is shown in
Figure 10. Here, the upper panel shows the Galactocentric rotational
velocity as a function of metallicity for the Extended Sample. The gray
dots represent the stars in the sample with [C/Fe] $< +0.7$, while the red
dots denote the stars with enhanced carbonicity, [C/Fe] $> +0.7$. It is
worth noting that the stars with [C/Fe] $>$ +0.7 are \emph{almost all}
located in the halo components of the Galaxy, with few exceptions. This is
perhaps not surprising, as the halo components of the Milky Way are very
metal poor. The middle panel of Figure 10 shows the Galactocentric
rotational velocity, \vphi\ , as a function of [C/Fe]. This plot shows no
evidence of correlation between V$_{\phi}$ and [C/Fe]. Finally, the lower
panel of Figure 10 shows [Fe/H] as a function of [C/Fe]. Here, there is
clear evidence of a correlation between [C/Fe] and [Fe/H] -- the
carbonicity increases as the metallicity decreases. A similar trend was
already noticed in past works, such as Rossi et al. (2005) and Lucatello et
al. (2006). We consider this result in more detail below.

\subsection{CEMP Fractions in the Low Metallicity Regime: Global Behavior}

The left panel of Figure 11 shows the fraction of CEMP stars in the
Extended Sample, as a function of [Fe/H], in the metal-poor regime ([Fe/H]
$<$ $-$1.5), and at vertical distance Z$_{\rm max}$ $>$ 5 kpc (chosen to
avoid thick disk and MWTD contamination). Here, each bin in metallicity
corresponds to an interval of $\Delta$[Fe/H] = 0.2 dex, with the exception
of the last bin, where stars are selected in the range [Fe/H] $<$ $-$2.6.
The fractions of CEMP stars in each bin are obtained by selecting objects
with [C/Fe] $>$ +0.7, and application of Eqn. 2. Errors on the CEMP
fraction are evaluated through the jackknife approach. This technique is
similar to bootstrapping, but instead of sampling with replacement, it
recomputes the statistical estimate leaving out one observation at a time
from the sample (Wall \& Jenkins 2003). The dash-dotted line is a
second-order polynomial fit to the data.

The increase of the fraction of CEMP stars with declining metallicity
pertains to the global behavior of the data in the metal-poor regime. If
metallicity alone is the driver of the carbon-enhancement phenomenon, one
might wonder if the strong increase in the fraction of CEMP stars at [Fe/H]
$< -$2.0 could be due to the increasing importance of the outer-halo
component, which has a peak of metallicity at [Fe/H] $\sim$ $-$2.2, and a
long tail extending to lower metallicity.

We have evaluated the as-observed fractions of carbon-rich stars in our
Extended Sample, for Z$_{\rm max}$ $>$ 5 kpc, over several bins in
metallicity. With the adopted definition of carbon-rich stars ([C/Fe] $>$
+0.7), and following Eqn. 2, we find that 2\%, 7\%, and 20\% of stars in
the intervals $-1.5 <$ [Fe/H] $< -0.5$, $-2.5 <$ [Fe/H] $< -1.5$, and
[Fe/H] $< -$2.5, are carbon-rich, respectively. For ease of comparison with
previous determinations of the CEMP fractions in the halo, we note that the
global fraction of CEMP stars in the halo system with \zmax\ $>$ 5 kpc and
[Fe/H] $< -1.5$ is 8\%, for [Fe/H] $< -2.0$ it is 12\%, and for \feh $<
-2.5$ it is 20\%.

The right panel of Figure 11 shows the mean carbonicity,
$\langle$[C/Fe]$\rangle$, as a function of [Fe/H]. Obviously, it is a
strong function of metallicity, although there may be some sign of it
leveling off at the lowest metallicities. Larger samples, in particular of
lower metallicity CEMP stars, are required to be certain.

Figure 12 shows the fraction of CEMP stars in the Extended Sample, as a
function of distance from the Galactic plane, $|$Z$|$, for stars satisfying
$-$2.0 $<$ [Fe/H] $< -$1.5 and [Fe/H] $< -$2.0, respectively. The intervals
for the cuts on height above or below the plane have widths of
$\Delta$$|$Z$|$ = 4 kpc, and the CEMP star fractions in each bin are
obtained by applying the same criteria and definitions as used for the left
panel of Figure 11. In the range of metallicity $-$2.0 $<$ [Fe/H] $< -$1.5,
significant contamination of the sample from thick-disk stars is not
expected, while the MWTD could be still present (with metallicity peak
around [Fe/H] $\sim -$1.3), but only in the region close to the Galactic
plane (0 kpc $<$ $|$Z$|$ < 4 kpc); the MWTD would not be expected to
contribute for the metallicity range [Fe/H] $< -$2.0 even close to the
plane. Inspection of this figure indicates a clear dependence of the CEMP
star fraction on distance from the Galactic plane. Close to the plane, this
dependence is due to the combined presence of the possible MWTD and
inner-halo populations, while at $|$Z$|$ $>$ 4 kpc, the observed fractions
must essentially pertain to the halo system alone. Far from the Galactic
plane, the observed increase of the CEMP star fractions with $|$Z$|$ would
be difficult to understand if the halo system comprises a single
population. In such a case, one might expect the CEMP star fractions to be
roughly constant, as a function of $|$Z$|$, for any given cut in
metallicity. This is clearly {\it not} what the data show.

Interestingly, we note that, for the same intervals in $|$Z$|$, the mean
carbonicity remains approximately contstant, at a value
$\langle$[C/Fe]$\rangle$ $\sim +1.0$ for $-$2.0 $<$ [Fe/H] $< -$1.5 and
$\langle$[C/Fe]$\rangle$ $\sim +1.5$ for \feh $< -2.0$. Of course, the
contribution from the inner- and outer-halo components is shifting as one
progresses from low to high $|$Z$|$, and this may be smoothing out any real
variations associated with the individual components. We return to this
question below.

A similar trend of increasing CEMP star fraction with height above the
Galactic plane was previously suggested by Frebel et al. (2006), based on a
much smaller sample of stars from the Hamburg/ESO survey (Wisotzki et al.
2000; Christlieb 2003), and proportionately larger error bars. Our expanded
data set now clearly indicates the existence of a strong spatial variation
of the CEMP star frequency {\it within the halo system}, and suggests that
the observed carbon enhancement is unlikely to be purely driven by
metallicity alone. Rather, it points to a real difference in CEMP star
fractions associated with the inner- and outer-halo populations, and opens
the possibility for different carbon-production mechanisms and/or different
chemical-evolution histories within their progenitors. We return to this
question below, after considering a method to probabilistically classify
individual stars as likely inner- or outer-halo members.

\subsection{The Extreme Deconvolution Technique}

Inference of the distribution function of an observable given only a
finite, noisy set of measurements of that distribution is a problem of
significant interest in many areas of science, and in astronomy in
particular. The observed distribution of a parameter is just the starting
point, but what is desired is knowledge of the distribution that we would
have in the case of very small uncertainties of the data and with all of
the dimensions of the parameter measured; in other words, the closest
representation of the underlying distribution. Usually, the data never have
these two properties, and it is then challenging to find the underlying
distribution without taking into account the uncertainty of the data (Bovy,
Hogg, \& Roweis, 2011; hereafter BHR11). The Extreme Deconvolution (XD)
technique of BHR11 confronts all of these issues, and provides an accurate
description of the underlying distribution of a
\emph{d}-dimensional quantity by taking into account the potentially
large and heterogeneous observational uncertainties, as well as missing
dimensions.

The BHR11 paper generalized the well known mixtures-of-Gaussians
density-estimation method to the case of noisy, heterogeneous, and
incomplete data. In this method, the underlying distribution of a
quantity {\bf v} is modeled as a sum of $K$ Gaussian distributions
\begin{eqnarray}
p(\textbf{v}) = \sum_{j=1}^K \alpha_{j}N(\textbf{v|m$_{j}$},\textbf{V$_{j}$})\,,
\end{eqnarray}
where the function N(\textbf{v|m$_{j}$},\textbf{V$_{j}$}) is the
\emph{d}-dimensional Gaussian distribution with mean \textbf{m} and
variance tensor \textbf{V} and $\alpha_{j}$ are the amplitudes,
normalized to sum to unity (all of these parameters are grouped
together as $\theta$ in what follows). The data {\bf w$_i$} are
assumed to be noisy samples from this distribution
\begin{eqnarray}
{\bf w_{i} = v_{i}} + \mathrm{noise}\,,
\end{eqnarray}
where the noise is drawn from a normal distribution with zero mean and
known covariance matrix {\bf S$_{i}$}. Here and in what follows, we
ignore the projection matrices {\bf R$_i$} of BHR11, since the data we
will apply this technique to are complete.

The likelihood of the model for each data point is given by the model
convolved with the uncertainty distribution of that data point. Since
a Gaussian distribution convolved with another Gaussian distribution
is again a Gaussian, the likelihood for each data point is a sum of
Gaussian distributions
\begin{eqnarray}
p(\textbf{w$_{i}$}|\theta) = \sum_{j=1}^K \alpha_{j}N(\textbf{w$_{i}$}|\textbf{m$_{j}$},\textbf{T$_{ij}$})\,
\end{eqnarray}
where
\begin{eqnarray}
\textbf{T$_{ij}$} = \textbf{V$_{j}$} + \textbf{S$_{i}$}\,.
\end{eqnarray}
The objective function is the total likelihood, obtained by simply
multiplying the individual likelihoods together for the various data
points
\begin{eqnarray}
\ln \mathcal{L} = \sum_{i}\ln p(\textbf{w$_{i}$}|\theta) = \sum_{i}\ln\sum_{j=1}^K \alpha_{j}N(\textbf{w$_{i}$}|\textbf{m$_{j}$},\textbf{T$_{ij}$}).
\end{eqnarray}

The optimization of this objective function provides the maximum likelihood
estimate of the distribution, or its parameters. In BHR11, the authors developed
a fast and robust algorithm to optimize the likelihood, based on an adaptation
of the expectation-maximization algorithm\footnote{Code implementing this
algorithm is available at \url{http://code.google.com/p/extreme-deconvolution/}\,.}
(Dempster et al. 1977).

The XD technique provides the best-fit values of the amplitude, mean,
and standard deviation of each Gaussian component, as well as the
so-called {\it posterior probability} that the observed data point
\textbf{w$_{i}$} is drawn from the component \emph{j}. The posterior
probability is given by
\begin{eqnarray}
p_{ij} = \frac{\alpha_{j}N(\textbf{w$_{i}$}|\textbf{m$_{j}$},\textbf{T$_{ij}$})}{\sum_{k}\alpha_{k}N(\textbf{w$_{i}$}|\textbf{m$_{k}$},\textbf{T$_{ik}$})}\,,
\end{eqnarray}
(see BHR11 for the derivation of this formula). The posterior
probability is a powerful statistical tool to perform probabilistic
assignments of stars to a Gaussian component in the model
distribution. In Galactic studies, the Gaussian component could
represent a primary structural component such as a disk or halo, a moving group,
or a spatial or velocity overdensity.

In many respects the XD technique described above is similar to the maximum
likelihood technique adopted in C10, but is more general, because it takes
into account the uncertainties of the measurements and provides the
membership probabilities.

\subsection{Application to the SDSS/SEGUE DR7 Calibration Stars}

The basic parameters for application of the XD approach for the analysis at
hand are the Galactocentric rotational velocity, V$_{\phi}$, the maximum
vertical distance, Z$_{\rm max}$, and the metallicity, [Fe/H]. We employ
the XD technique to determine the underlying distribution of the rotational
velocity of halo stars and to determine the posterior membership
probabilities for each star. In practice, the entries for the XD algorithm
are the Galactocentric rotational velocity and its uncertainty for each
star, V$_{\phi,i}$ and $\varepsilon_{V_{\phi},i}$, respectively\footnote
{Note that errors in the rotational velocity depend in turn on errors in
the distances and proper motions; these are carried forward into the
analysis automatically.}. These two parameters, together with all of the other
kinematic and orbital quantities, are derived using the same procedures
employed by C07 and C10 (but with revised distances for the reassigned
main-sequence turnoff stars, as discussed above).

Our plan is to make use of a sample in which the constraint on the distance
from the Sun, $d$, and on the projected Galactic distance, $R$, are relaxed
(in order to take advantage of the larger numbers of CEMP stars in the
Extended Sample). As a consequence, the data set becomes noisier then the
Local Sample, because uncertainties on the transverse velocities increase
with the distance $d$. In the terminology of the XD approach, the analysis
of the Extended Sample falls into the case where all the observables are
known, but some of the derived parameters have large uncertainties.

The Galactic halo is assumed to be a two-component structure, comprising the inner
and the outer halo, as discussed in C07 and C10.
Thus, the general expression for the likelihood takes the form:
\begin{eqnarray}
\ln \mathcal{L} = \displaystyle\sum_{i} \ln [\alpha_{in}\cdot N^{i}_{in} + \alpha_{out}\cdot N^{i}_{out}]\,,
\end{eqnarray}
where $\alpha_{in}$ and $\alpha_{out}$ are the amplitude of the
inner halo and outer halo, respectively. The velocity distributions are
assumed to be Gaussian, thus
\begin{eqnarray}
N^i_{in/out} = N(V_{\phi,i} | V_{\phi,in/out},\sigma^2_{\phi,in/out}+\varepsilon^2_{V_{\phi},i})\,,
\end{eqnarray}

The membership probabilities then take the form:

\begin{eqnarray}
p_{i, in} = \frac{\alpha_{in}N_{in}^{i}}{\alpha_{in}N_{in}^{i} + \alpha_{out}N_{out}^{i}}
\end{eqnarray}

and

\begin{eqnarray}
p_{i, out} = \frac{\alpha_{out}N_{out}^{i}}{\alpha_{in}N_{in}^{i} + \alpha_{out}N_{out}^{i}}
\end{eqnarray}

for the \emph{i}th star.\\

\subsubsection{Extreme Deconvolution Results for the Extended Sample}

In this section the posterior probabilities, derived through the
application of the XD, are obtained for the Extended Sample ($d <$ 10 kpc,
no constraints on $R$), selected in the range of metallicity [Fe/H] $<
-2.0$, and with vertical distance Z$_{\rm max}$ $>$ 5 kpc. The results are
shown in Figure 13. The upper panel of this figure shows the distribution
of the observed rotation velocity V$_{\phi}$ (black histogram), and the
green and red curves represent the resulting Gaussian velocity
distributions for the inner- and outer-halo components. The middle panel
shows the derived velocity distribution for the two components, obtained by
weighting each star with its membership probability, and normalized such
that the total area corresponds to unity. The black histogram is the
observed distribution as in the top panel, but normalized to unity as well.
In the lower panel, the posterior probability as a function of the
rotational velocity, V$_{\phi}$, is shown. This probability has been
obtained using Eqns. 11 and 12. For each star in the sample, the XD
technique assigns two probability values, the first associated with the
inner halo, and the second related to the outer halo. Since in our model a
star is required to belong to one or the other component (there are no
orphans allowed), and the summed inner- and outer-halo posterior
probabilities are forced to unity at each point, the two curves simply
complement one another.

The values obtained for the mean rotational velocity and its dispersion
are, for the inner halo, V$_{\phi}$ = 56 $\pm$ 11 km s$^{-1}$, and
$\sigma_{V_{\phi}}$ = 93 $\pm$ 35 km s$^{-1}$; for the outer halo,
V$_{\phi}$ = $-$141 $\pm$ 31 km s$^{-1}$, and $\sigma_{V_{\phi}}$ = 138
$\pm$ 58 km s$^{-1}$. The errors on the velocity and its dispersion have
been evaluated by employing the jackknife method. A K-S test of the null
hypothesis that the velocity distributions of stars belonging to the inner-
and outer-halo components could be drawn from the same parent population is
rejected at a high level of statistical significance (\emph{p} $< 0.001$).

\subsection{Contrast of CEMP Stellar Fractions in the Two Halo Components}

We now consider the distribution of carbonicity ([C/Fe]) for the Extended
Sample with metallicity [Fe/H] $< -$2.0 and at Z$_{\rm max}$ $>$ 5 kpc.
These selections ensure the presence of essentially all halo stars in the
sample, with little or no contamination expected from the thick disk or
MWTD. The top panel of Figure 14 shows the CarDF for the Extended
Sample\footnote{Note that the number of stars shown in Figure 14 differs
from the numbers shown in Figure 13, as Figure 14 includes only stars with
detected CH G-band features.}. There are two clear peaks that emerge, one
at [C/Fe] $\sim +0.3$ to $+0.5$, and the second softer peak around [C/Fe]
$\sim +1.8$, along with a tail extending towards higher values of
carbonicity. The bottom panel of Figure 14 shows the weighted CarDFs for
the inner- and outer-halo components. The underlying distribution of each
population has been obtained by weighting the values of the CarDF with the
membership probabilities of each star. As usual, the green distribution
denotes the inner halo, while the red distribution represents the outer
halo. A K-S test of the null hypothesis that the CarDF of stars belonging
to the inner- and outer-halo components could be drawn from the same parent
population is rejected at high statistical significance (\emph{p} $<$
0.001), clearly indicating that the carbonicity distribution functions of
the inner- and outer-halo components are distinct.

The contrast of the CEMP star fractions in the inner- and outer-halo
components can be investigated by applying the XD analysis to subsamples of
stars in restricted ranges of metallicity. The stars employed are those
belonging to the Extended Sample, selected in the low metallicity range
([Fe/H] $<$ $-$1.5) and with Z$_{\rm max}$ $>$ 5 kpc.

A first attempt was made by selecting stars in intervals of metallicity
such that $\Delta$[Fe/H] = 0.2 dex, and applying the XD technique to the
rotational velocity distribution of each subsample. This experiment was not
successful, due to the low number of stars in each bin. A much better
result has been obtained by choosing a larger interval on metallicity,
$\Delta$[Fe/H] = 0.5 dex.

The left panel of Figure 15 shows the previously determined global trend of
the CEMP star fraction, as a function of metallicity (dashed curve),
overplotted with the derived inner- and outer-halo CEMP star fractions. The
blue filled stars represent the predicted values of the CEMP star fractions
for each bin of metallicity, i.e., $-$2.0 $<$ [Fe/H] $<-$ 1.5, $-$2.5 $<$
[Fe/H] $<-$ 2.0, and $-$3.0 $<$ [Fe/H] $<-$ 2.5 (hereafter bin1, bin2, and
bin3). The predicted values are derived using the global trend of CEMP star
fractions vs. [Fe/H] (second-order polynomial; Fig. 11), and the posterior
probability for each star in the subsamples associated with each bin of
metallicity, such that:

\begin{equation}
F^{k}_{CEMP_{pred}} = \frac{1}{\sum_{i}(p_{ij})}\cdot\sum_{i}(p_{ij}f^{k})
\end{equation}

\noindent where p$_{ij}$ is the posterior probability, derived by applying the XD
analysis to the subsample of stars selected in bin1, bin2, and bin3,
respectively. The parameter $f^{k}$ represents the CEMP star fractions
obtained in each bin of metallicity from application of the second order
polynomial (\emph{j} and \emph{k} denote the halo component and the bin of
metallicity, respectively).

Note that the expression in Eqn. 13 provides an estimate of the CEMP
fraction that is driven by the global trend of F$_{CEMP}$ as a function of
the metallicity (Fig. 11, left panel). Eqn. 13 makes use of the full shape
of the posterior probability, and thus, a considerable overlap between the
kinematic distributions of the two components is still present. Therefore,
we are not expecting to find a significant difference in the predicted
values for the two components in each bin of [Fe/H].

The predicted value of CEMP star fractions in bin1 is F$_{CEMP_{pred,in,
1}}$ = (5.8 $\pm$ 0.03)\%, F$_{CEMP_{pred,out, 1}}$ = (5.9 $\pm$ 0.03)\%,
for the inner halo and outer halo, respectively. The same calculations for
bin2 provide F$_{CEMP_{pred,in, 2}}$ = (11.9 $\pm$ 0.1)\% and
F$_{CEMP_{pred,out, 2}}$ = (12.0 $\pm$ 0.1)\%. For the the third bin we
find F$_{CEMP_{pred,in, 3}}$ = (22.9 $\pm$ 0.6)\% and F$_{CEMP_{pred,out,
3}}$ = (22.7 $\pm$ 0.6)\%. The error on each fraction is evaluated by
employing the jackknife method. These fractions are sufficiently close to
one another that only a single set of blue stars is shown in the left panel
of Figure 15 to represent both results.

Not surprisingly, the above predicted values of the CEMP star fractions in
the inner and outer halo are very similar in all bins of metallicity. The
presence of a contrast in the CEMP star fraction for the two components can
be better investigated by reducing the overlap in the kinematic
distributions. This can be done by employing the
\emph{hard-cut-in-probability} method. For each halo component, the lower
limit of the membership probability is chosen to be $p_{lim}$ = 0.7. This
limit is set reasonably high to reduce the contamination between the halo
components. We have selected two subsamples of stars such that p$_{i,inner}
>$ $p_{lim}$, and p$_{i,outer} >$ $p_{lim}$ (\emph{i} denotes the star),
and for each subsample, the CEMP star fraction is evaluated following Eqn.
2.

The result of this exercise is shown in the left panel of Figure 15. Here,
the green (inner) and the red (outer) filled circles denote the values of
the CEMP star fractions obtained by employing the hard-cut-in-probability
method. The observed CEMP star fractions in bin1 are F$_{CEMP_{obs, in,
1}}$ = (5.2 $\pm$ 0.6)\% and F$_{CEMP_{obs,out, 1}}$ = (7.7 $\pm$ 1.0)\%,
for the inner and outer halo, respectively. In the case of bin2, we find
F$_{CEMP_{obs, in, 2}}$ = (9.4 $\pm$ 0.8)\% and F$_{CEMP_{obs,out, 2}}$ =
(20.3 $\pm$ 2.3)\%. Finally, in the third bin, the observed fractions are
F$_{CEMP_{obs,in, 3}}$ = (20.5 $\pm$ 2.4)\% and F$_{CEMP_{obs, out, 3}}$ =
(30.4 $\pm$ 4.8)\%.

In the highest metallicity bin, $-2.0 <$ [Fe/H] $< -1.5$, the CEMP star
fractions for the two halo components are close to the expected value
($\sim$ 6\%), with overlapping error bars. Note that the inner halo is the
dominant component in this range of metallicity ([Fe/H]$_{peak,in}$ =
$-$1.6), and we were not expecting to find a high contrast in the CEMP star
fractions. At lower metallicity, $-2.5 <$ [Fe/H] $< -2.0$, where the
dominant component is the outer halo ([Fe/H]$_{peak,out} = -2.2$), there is
evidence of a significant contrast in CEMP fractions between the inner- and
outer-halo components, such that F$_{CEMP_{obs, out, 2}}$ $\sim$
2$\times$F$_{CEMP_{obs, in, 2}}$. This contrast is confirmed also in the
lowest bin of metallicity, $-3.0 <$ [Fe/H] $< -2.5$, even though it is less
remarkable, F$_{CEMP_{obs, out, 3}}$ $\sim$ 1.5$\times$F$_{CEMP_{obs, in,
3}}$, and has a much larger error bar, due to the smaller numbers of stars
involved.

It is important to note that most of the stars in the lowest metallicity
bin belong to the category for which only an upper limit of [C/Fe] has been
provided (classified as subsample L, see Section 2.4). A significant number
of stars in the lowest metallicity bin have high \teff\ ($\sim$ 65\% above
6250~K), and thus only quite high values of [C/Fe] are expected to be
detected. Thus, it is our expectation that the quoted frequencies for CEMP
star fractions, at least at low metallicity, represent lower limits.
Morever, the SDSS/SEGUE DR7 calibration-star sample has a limited number of
stars with [Fe/H] $< -2.5$. The extremely metal-poor regime will be further
investigated in a future paper using the much larger sample of such stars
in the SDSS/SEGUE DR8 release (Aihara et al. 2011).

Interestingly, the right panel of Figure 15 shows that the mean
carbonicity, $\langle$[C/Fe]$\rangle$, remains similar for both the inner-
and outer-halo components as a function of declining metallicity. There is
a hint of a split emerging for the lowest metallicity bin, $-3.0 <$ [Fe/H]
$< -2.5$, with the outer-halo $\langle$[C/Fe]$\rangle$ being slightly
higher than that of the inner-halo value. However, the error bars overlap,
indicating that the sample size is simply too small to be certain.

\section{Summary and Discussion}

We have analyzed the SDSS/SEGUE DR7 calibration stars in order to explore
the global properties of the carbonicity distribution function ([C/Fe];
CarDF) of the halo components of the Milky Way, and the possible contrast
of the CEMP star fractions and mean carbonicity between the inner halo and
the outer halo. Carbon-to-iron abundance ratios (or limits) have been
obtained for 31187 stars, based on matches to a grid of synthetic spectra
to the CH G-band at 4305 {\AA}, with an external error for [C/Fe] on the
order of 0.25 dex. Kinematic and orbital parameters were derived employing
the methodologies described in C10. We have considered the nature of the
samples of stars considered CEMP ([C/Fe] $> +0.7$) and non-CEMP ([C/Fe] $<$
+0.7), and shown that, at low metallicity ([Fe/H] $< -1.5$), their
distributions on distance and rotational velocity differ significantly from
one another. For the stars with a detectable CH G-band, a deconvolution of
the inner- and outer-halo components, along with a derivation of the
membership probabilities, has been obtained by applying the Extreme
Deconvolution (XD) analysis to the Extended Sample of calibration stars,
selected in the low metallicity range ([Fe/H] $< -$1.5) and with Z$_{max}
>$ 5 kpc. Contrasts of the CEMP star fractions and mean carbonicity have
been obtained by employing the hard-cut-in-probability method to subsamples
of stars selected in several bins of metallicity.

\subsection{Summary of Main Results}

Our main results are the following:

\begin{itemize}

\item The as-observed distribution of [C/Fe], which we refer to as the
carbonicity distribution function (CarDF), has been derived at various
intervals on distance above or below the Galactic plane. At $|$Z$|$ $>$ 5
kpc, where we expect to see the beginning of the transition from inner-halo
dominance to outer-halo dominance among halo stars, the CarDF exhibits a
strong peak at [C/Fe] $\sim$ +0.3, and a long tail towards high values of
carbonicity, up to [C/Fe] = +3.0. The fraction of CEMP stars (defined here
to mean [C/Fe] $>$ $+$0.7) for the subsample at $|$Z$|$ $>$ 9 kpc (where
the dominant component is the outer halo) is 20\%, in agreement with
several previous estimates for stars at similarly very low metallicities,
[Fe/H] $<-$2.0.

\item At low metallicities ([Fe/H] $< -1.5$), the distribution of derived
distances and rotational velocities for the CEMP and non-CEMP stars differ
significantly from one another.

\item \emph{Almost all} of the CEMP stars are located in the halo components
of the Milky Way.

\item We observe a significant increase of the fraction of CEMP
stars with declining metallicity in the halo system. Following Eqn. 2, we
find that 2\%, 7\%, and 20\% of stars in the metallicity intervals $-1.5 <$
[Fe/H] $< -0.5$, $-2.5 <$ [Fe/H] $< -1.5$, and [Fe/H] $< -$2.5, are
carbon-rich, respectively. For ease of comparison with previous estimates,
the global frequency of CEMP stars in the halo system for
\feh\ $< -1.5$ is 8\%; for \feh\ $< -2.0$ it is 12\%; for \feh\ $<-2.5$ it is 20\%.
For reasons described above, we believe these fractions should be
considered lower limits.

\item In the low-metallicity regime ([Fe/H] $< -$1.5), and at vertical
distances Z$_{max} >$ 5 kpc, a continued increase of the fraction of CEMP
stars with declining metallicity is found (Fig. 11, left panel). A
second-order polynomial provides a good fit to the CEMP star fraction as a
function of [Fe/H].

\item In the same metallicity regime the mean carbonicity,
$\langle$[C/Fe]$\rangle$, increases with declining metallicity (Fig. 11,
right panel).

\item In the low-metallicity regime, for both $-2.0 <$ [Fe/H] $< -$1.5
and [Fe/H] $< -$2.0, we find a clear increase in the CEMP star fraction
with distance from the Galactic plane, $|$Z$|$ (Fig. 12). At these low
metallicities, significant contamination from the thick disk and MWTD
populations is unlikely, and would only have a small effect for
the bin with 0 kpc $<$ $|$Z$|$ $<$ 4 kpc and $-2.0 <$ [Fe/H] $< -$1.5; this
is an observed property of the halo system. The mean carbonicity remains
roughly constant as a function of $|$Z$|$, taking a value of
$\langle$[C/Fe]$\rangle$ $\sim +1.0$ for $-2.0 <$ [Fe/H] $< -$1.5, and
$\langle$[C/Fe]$\rangle$ $\sim +1.5$ for \feh\ $< -2.0$.

\item The Extreme Deconvolution technique of BHR11 has
been applied to subsamples of stars selected in three bins of metallicity,
$-$2.0 $<$ [Fe/H] $<$ $-$1.5, $-$2.5 $<$ [Fe/H] $<$ $-$2.0, and $-$3.0 $<$
[Fe/H] $<$ $-$2.5, and with Z$_{\rm max}$ $>$ 5 kpc. We have successfully
decomposed the inner and outer halo in all three bins, and obtained the
inner- and outer-halo membership probabilities for each star.

\item At low metallicity, $-2.0 <$ [Fe/H] $< -1.5$, where the dominant component
is the inner halo ([Fe/H]$_{peak,in} = -1.6$), there is no significant
difference in the CEMP star fraction btween the inner and outer halo. In
contrast, at very low metallicity, $-2.5 <$ [Fe/H] $< -2.0$, where the
dominant component is the outer halo ([Fe/H]$_{peak,out} = -2.2$), there is
evidence for a significant difference in the CEMP star fraction between the
inner and the outer halo (Figure 15, left panel), such that F$_{CEMP_{obs,
out, 2}}$ $\sim$ 2$\times$F$_{CEMP_{obs, in, 2}}$. This difference is also
confirmed in the lowest metallicity bin, $-3.0 <$ [Fe/H] $< -2.5$, even
though it is less remarkable, F$_{CEMP_{obs, out, 3}}$ $\sim$
1.5$\times$F$_{CEMP_{obs, in, 3}}$, and has a larger error bar. We conclude
that the difference in CEMP frequency at very low metallicity is not driven
by metallicity itself, but rather, by the stellar populations present.

\item The mean carbonicity, $\langle$[C/Fe]$\rangle$, remains similar for both the
deconvolved inner- and outer-halo components as a function of declining
metallicity. Larger samples of low-metallicity CEMP stars are required to
see if a split in mean carbonicity emerges.

\end{itemize}

\subsection{Implications for Galaxy Formation}

A possible scenario for the formation of the inner- and the outer-halo
components has been described in C07. The fact that the outer-halo
component of the Milky Way exhibits a net retrograde rotation (and a
different distribution of overall orbital properties), clearly indicates
that the formation of the outer halo is distinct from that of the inner
halo and the disk components. In C07, it was suggested that the outer-halo
component formed, not through a dissipational, angular-momentum conserving
contraction, but rather through dissipationless chaotic merging of smaller
subsystems within a pre-existing dark matter halo. These subsystems would
be expected to be of much lower mass, and subject to tidal disruption in
the outer part of a dark matter halo, before they fall farther into the
inner part. As candidate (surviving) counterparts for such subsystems, one
might consider the low luminosity dwarf spheroidal galaxies surrounding the
Milky Way. In the case of the inner halo, C07 argued that the low-mass
sub-Galactic fragments, formed at an early stage, rapidly merge into several
more massive clumps, which themselves eventually dissipationally merge
(owing of the presence of gas that had yet to form stars). The essentially
radial merger of the few resulting massive clumps gives rise to the
dominance of high-eccentricity orbits for stars that we assign to
membership in the inner halo. Star formation within these massive clumps
(both pre- and post-merger) would drive the mean metallicity to higher
abundances. This would be followed by a stage of adiabatic compression
(flattening) of the inner-halo component owing to the growth of a massive
disk, along with the continued accretion of the gas onto the Galaxy. This
general picture is supported, at least qualitatively, by the most recent
numerical simulations of the formation and evolution of large disk galaxies
that include prescriptions to handle the presence of gas, and which also
include approximate star-formation schemes and follow the evolving
chemistry (e.g., Zolotov et al. 2010; Font et al. 2011; McCarthy et al.
2011; Tissera et al. 2011).

The large fractions of CEMP stars in both halo components indicates that
significant amounts of carbon were produced in the early stages of chemical
evolution in the universe. The observed contrast of CEMP star fractions
between the inner halo and outer halo strengthen the picture that the halo
components had different origins, and supports a scenario in which the
outer-halo component has been assembled by the accretion of small
subsystems, as discussed below. In this regard, it is interesting that the
MDF of the inner halo (peak metallicity at [Fe/H] $\sim -$ 1.6) is in the
metallicity regime associated with the CEMP-s stars, which are primarily
found with [Fe/H] $> -2.5$, while the MDF of the outer halo (metallicity
peak at [Fe/H] $\sim$ $-$2.2) might be associated with the metallicity
regime of the CEMP-no stars, which are primarily found with [Fe/H] $< -2.5$
(Aoki et al. 2007b).

The fact that, in the metal-poor regime, the outer halo exhibits a fraction
of CEMP stars that is larger than the inner halo (CEMP$_{outer}$ $\sim$
2$\cdot$CEMP$_{inner}$), suggests that multiple sources of carbon, besides
the nucleosynthesis of AGB stars in binary systems, were present in the
pristine environment of the outer-halo progenitors (lower mass sub-halos).
These sources could be the fast massive rotators and/or the faint
supernovae mentioned in the Introduction. Karlsson (2006) also explored the
possibility that primordial gas was pre-enriched in heavy metals by less
massive SNe (13 $<$ M/M$_{\odot}$ $<$ 30), whose ejecta underwent
substantial mixing and fallback, while the C (and N, O) originated from
massive rotating stars with M $\geq$ 40 M$_{\odot}$. If the CEMP stars in
the outer halo are predominantly CEMP-no stars (which has yet to be
established), it might suggest that non-AGB-related carbon production took
place in the primordial mini-halos. The predominance of CEMP-s stars in the
inner halo, if found, would suggest that the dominant source of carbon was
the nucleosynthesis by AGB stars in binary systems. This would place
important constraints on the primordial IMF of the sub-systems responsible
for the formation of the two halo components.

Recent efforts to model, from the population synthesis standpoint, the
fractions of observed CEMP stars in the halo have not managed to reproduce
results as high as 15-20\% for metallicities [Fe/H] $< -2.0$ (e.g., Izzard
et al. 2009; Pols et al. 2010). However, these predictions are
based solely on carbon production by AGB stars. While such calculations may
prove meaningful for the inner-halo population, they may not be telling the
full story for carbon production associated with the progenitors of the
outer-halo population. Indeed, the observed increase of the CEMP star
fraction with $|$Z$|$ (in particular, far from the Galactic plane) we have
found would be difficult to understand if the halo system represents a
single population. In any event, the lower CEMP fraction in the inner halo
may relieve some of the tension with current model predictions.

The fact that the CEMP star fraction exhibits a clear increase with $|$Z$|$
suggests that the relative numbers of CEMP stars in a stellar population is
not driven by metallicity alone. The proposed coupling of the cosmic
microwave background to the initial mass function (CMB-IMF hypothesis;
Larson 1998, 2005; Tumlinson 2007) is one mechanism for imposing a temporal
dependence on the IMF. This effect, coupled with chemical evolution models,
predicts that the CEMP fraction would be expected to increase as the
metallicity decreases, but with similar metallicity regimes forming carbon
according to the expected yields of the predominant mass range available at
that time. In the hierarchical context of galaxy formation, star-forming
regions are spatially segregated, and their chemical evolution can proceed
at different rates, with stars at the same metallicity forming at different
times. Depending on the source(s) of carbon, this trend could lead to a
spatial variation of the CEMP fraction at the same metallicity, increasing
in older populations and decreasing in younger ones.

The larger fraction of CEMP stars associated with the outer-halo component
could be due to the nature of the progenitor low-mass mini-halos (mass,
density, etc.) in which they formed. A natural place to look for surviving
examples are the ultra-faint dwarf spheroidal galaxies surrounding the
Milky Way discovered in the course of the SDSS (Willman et al. 2005;
Belokurov et al. 2006a,b; Zucker et al. 2006, and many others since). Early
hints of the possible association of CEMP stars with the ultra-faint
galaxies came from the recognition that a (serendipitously)
spectroscopically targeted star from SDSS in the direction of the Canes
Venatici ultra-faint dwarf, SDSS~J1327+3335, is a carbon-rich giant with a
radial velocity and inferred distance commensurate with this very
metal-poor satellite galaxy (Zucker et al. 2006). The existence of CEMP
stars in ultra-faint dwarf spheroidal galaxies has now been definitively
established (Norris et al. 2010a,b; Lai et al. 2011). Norris et al. 2010b
reported the discovery of an extremely carbon-rich red giant, Segue 1-7, in
the Segue 1 system. This star exhibits a metallicity [Fe/H] = $-$3.5,
carbonicity [C/Fe] $= +2.3$, and a low barium abundance ratio ([Ba/Fe] $<
-$1.0), which place it in the CEMP-no category. This discovery is
consistent with the idea that the CEMP-no stars may indeed be associated
with the ultra-faint dwarf spheroidal galaxies; further similar
investigations should prove of great interest.

\acknowledgments

D.C. gratefully acknowledges funding from RSAA ANU to pursue her research.
T.C.B. and Y.S.L. acknowledge partial funding of this work from grants PHY
02-16783 and PHY 08-22648: Physics Frontier Center/Joint Institute for Nuclear
Astrophysics (JINA), awarded by the U.S. National Science Foundation. J.B. was
partially supported by NASA (grant NNX08AJ48G) and the NSF (grant AST 09-08357).
Studies at ANU of the most metal-poor populations of the Milky Way are supported
by Australian Research Council grants DP0663562 and DP0984924, which are
gratefully acknowledged by J.~E.~N. and D.~C.
\\





{\it Facilities:} \facility{SDSS}.

{}
\clearpage

\begin{deluxetable}{cccccccccrr}
\tablewidth{0pt}
\tabletypesize{\scriptsize}
\tablenum{1}
\tablecaption{Comparison with High-Resolution Spectroscopy: \\Atmospheric
Parameters and [C/Fe] Determinations}
\tablehead{
\colhead{} &
\colhead{} &
\colhead{} &
\colhead{\teff\ } &
\colhead{\teff\ } &
\colhead{\logg\ } &
\colhead{\logg\ } &
\colhead{[Fe/H]} &
\colhead{[Fe/H]} &
\colhead{[C/Fe]} &
\colhead{[C/Fe]} \\
\colhead{IAU Name} &
\colhead{PLATE-MJD-FIBER} &
\colhead{REF} &
\colhead{HIGH} &
\colhead{SSPP} &
\colhead{HIGH} &
\colhead{SSPP} &
\colhead{HIGH} &
\colhead{SSPP} &
\colhead{HIGH} &
\colhead{SSPP}
}
\startdata					
SDSS J000219.9+292851.8 & 2803-54368-459 & A11 & 6150  &  6184 & 4.00 & 3.24 & $-$3.26& $-$2.97 &$+$2.63 &$+$2.73  \\
SDSS J003602.2-104336.3 & \tablenotemark{a}{0654-52146-011} & A08 & 6500  &  6475 & 4.50 & 4.59 & $-$2.41& $-$2.49 &$+$2.50 &$+$2.21  \\
                        &                &     &       &       &      &      &        &         &        &         \\
SDSS J012617.9+060724.8 & 2314-53713-090 & A08 & 6600  &  6841 & 4.10 & 4.73 & $-$3.11& $-$2.80 &$+$2.92 &$+$2.67  \\
SDSS J012617.9+060724.8 & 2314-53713-090 & A11 & 6900  &  6841 & 4.00 & 4.73 & $-$3.01& $-$2.80 &$+$3.08 &$+$2.67  \\
                        &                &     &       &       &      &      &        &         &        &         \\
SDSS J025956.4+005713.3 & 1513-53741-338 & A11 & 4550  &  4537 & 5.00 & 4.31 & $-$3.31& $-$3.75 &$-$0.02 &$+$0.00  \\
SDSS J030839.3+050534.9 & 2335-53730-314 & A11 & 5950  &  5938 & 4.00 & 3.36 & $-$2.19& $-$2.41 &$+$2.36 &$+$2.31  \\
SDSS J035111.3+102643.2 & 2679-54368-543 & A11 & 5450  &  5542 & 3.60 & 3.14 & $-$3.18& $-$2.89 &$+$1.55 &$+$1.61  \\
SDSS J071105.4+670128.2 & 2337-53740-564 & A11 & 5350  &  5252 & 3.00 & 2.70 & $-$2.91& $-$2.91 &$+$1.98 &$+$2.16  \\
SDSS J072352.2+363757.2 & 2941-54507-222 & A11 & 5150  &  5105 & 2.20 & 2.33 & $-$3.32& $-$3.48 &$+$1.79 &$+$1.39  \\
SDSS J074104.2+670801.8 & 2939-54515-414 & A11 & 5200  &  5171 & 2.50 & 2.35 & $-$2.87& $-$2.92 &$+$0.74 &$+$0.84  \\
                        &                &     &       &       &      &      &        &         &        &         \\
SDSS J081754.9+264103.8 & \tablenotemark{a}{1266-52709-432} & A08 & 6300  &  6075 & 4.00 & 3.58 & $-$3.16& $-$3.03 & $<+$2.20&$<+$1.54 \\
SDSS J081754.9+264103.8 & \tablenotemark{a}{1266-52709-432} & A11 & 6050  &  6075 & 4.00 & 3.58 & $-$2.85& $-$3.03 & \dots  &$<+$1.54 \\
                        &                &     &       &       &      &      &        &         &        &         \\
SDSS J091243.7+021623.7 & \tablenotemark{a}{0471-51924-613} & A11 & 6150  &  6138 & 4.00 & 3.40 & $-$2.68& $-$2.76 &$+$2.05 &$+$2.33  \\
SDSS J091243.7+021623.7 & \tablenotemark{a}{0471-51924-613} & B10 & 6500  &  6138 & 4.50 & 3.40 & $-$2.50& $-$2.76 &$+$2.17 &$+$2.33  \\
                        &                &     &       &       &      &      &        &         &        &         \\
SDSS J092401.9+405928.7 & \tablenotemark{a}{0938-52708-608} & A08 & 6200  &  6201 & 4.00 & 4.53 & $-$2.51& $-$2.81 &$+$2.72 &$+$2.56  \\
SDSS J103649.9+121219.8 & \tablenotemark{a}{1600-53090-378} & B10 & 6000  &  5873 & 4.00 & 3.12 & $-$3.20& $-$3.31 &$+$1.47 &$+$1.94  \\
SDSS J124123.9-083725.5 & 2689-54149-292 & A11 & 5150  &  5108 & 2.50 & 2.45 & $-$2.73& $-$2.90 &$+$0.50 &$+$0.65  \\
SDSS J124204.4-033618.1 & 2897-54585-210 & A11 & 5150  &  5112 & 2.50 & 2.56 & $-$2.77& $-$3.02 &$+$0.64 &$+$0.77  \\
                        &                &     &       &       &      &      &        &         &        &         \\
SDSS J134913.5-022942.8 & \tablenotemark{a}{0913-52433-073} & A11 & 6200  &  6167 & 4.00 & 4.39 & $-$3.24& $-$3.16 &$+$3.01 &$+$2.61  \\
SDSS J134913.5-022942.8 & \tablenotemark{a}{0913-52433-073} & B10 & 6200  &  6167 & 4.00 & 4.39 & $-$3.00& $-$3.16 &$+$2.82 &$+$2.61  \\
                        &                &     &       &       &      &      &        &         &        &         \\
SDSS J161226.2+042146.6 & 2178-54629-546 & A11 & 5350  &  5365 & 3.30 & 2.42 & $-$2.86& $-$3.19 &$+$0.63 &$+$0.98  \\
SDSS J161313.5+530909.7 & 2176-54243-614 & A11 & 5350  &  5338 & 1.60 & 2.67 & $-$3.32& $-$2.89 &$+$2.09 &$+$1.64  \\
SDSS J164610.2+282422.2 & \tablenotemark{a}{1690-53475-323} & A11 & 6100  &  6125 & 4.00 & 3.45 & $-$3.05& $-$2.78 &$+$2.52 &$+$2.53  \\
SDSS J170339.6+283649.9 & 2808-54524-510 & A11 & 5100  &  5120 & 4.80 & 3.58 & $-$3.21& $-$3.32 &$+$0.28 &$+$0.43  \\
SDSS J170733.9+585059.7 & 0353-51703-195 & A08 & 6700  &  6567 & 4.20 & 3.47 & $-$2.52& $-$2.68 &$+$2.10 &$+$2.43  \\
SDSS J173417.9+431606.5 & 2799-54368-138 & A11 & 5200  &  5172 & 2.70 & 2.15 & $-$2.51& $-$3.18 &$+$1.78 &$+$2.43  \\
SDSS J204728.8+001553.8 & \tablenotemark{a}{0982-52466-480} & A08 & 6600  &  6324 & 4.50 & 3.77 & $-$2.05& $-$2.28 &$+$2.00 &$+$1.98

\enddata
\tablecomments{A08: Aoki et al. (2008); B10: Behara et al. (2010); A11: Aoki
et al. (2011, in prep.).  Note that the stars SDSS~J012617.9+060724.8 and
SDSS~J081754.9+264103.8 were reported on by both A08 and A11;
the stars SDSS~J091243.7+021623.7 and SDSS~J134913.5-022942.8 were reported on by both B10 and A11.}
\tablenotetext{a}{This star is a member of the calibration star sample
analyzed in this paper.}
\end{deluxetable}
\clearpage

\begin{deluxetable}{cccccc}
\tablewidth{0pt}
\tablenum{2}
\tablecaption{Numbers of Stars with Detected and Undetected\\ CH G-Bands and
Fractions of Carbon-rich Stars\\ as a Function of Metallicity}
\tablehead{
\colhead{Metallicity} &
\colhead{N$^{D}_{C-norm}$} &
\colhead{N$^{D}_{C-rich}$} &
\colhead{N$^{L}_{C-norm}$} &
\colhead{N$^{L}_{C-un}$} &
\colhead{F$_{C-rich}$} \\
\colhead{} &
\colhead{} &
\colhead{} &
\colhead{} &
\colhead{} &
\colhead{\%}
}
\startdata
  $-$1.5 $<$ [Fe/H] $\leq$ $-$1.0   & 9113 & 149 &  335 &   3 & 1.6  \\
  $-$2.0 $<$ [Fe/H] $\leq$ $-$1.5   & 4376 & 377 & 2651 & 226 & 5.1  \\
  $-$2.5 $<$ [Fe/H] $\leq$ $-$2.0   &  506 & 173 & 1387 & 458 & 8.4  \\
  $-$4.0 $<$ [Fe/H] $\leq$ $-$2.5   &   13 &  46 &  201 & 201 & 18.0
\enddata
\tablecomments{{\it C-un} indicates stars with unknown carbon status (see text).
Unlike the rest of the samples discussed in this paper,
this table includes numbers without regard as to whether they have acceptable
measured proper motions and radial velocities.}
\end{deluxetable}
\clearpage

\begin{figure}
\figurenum{1}
\begin{center}
\includegraphics{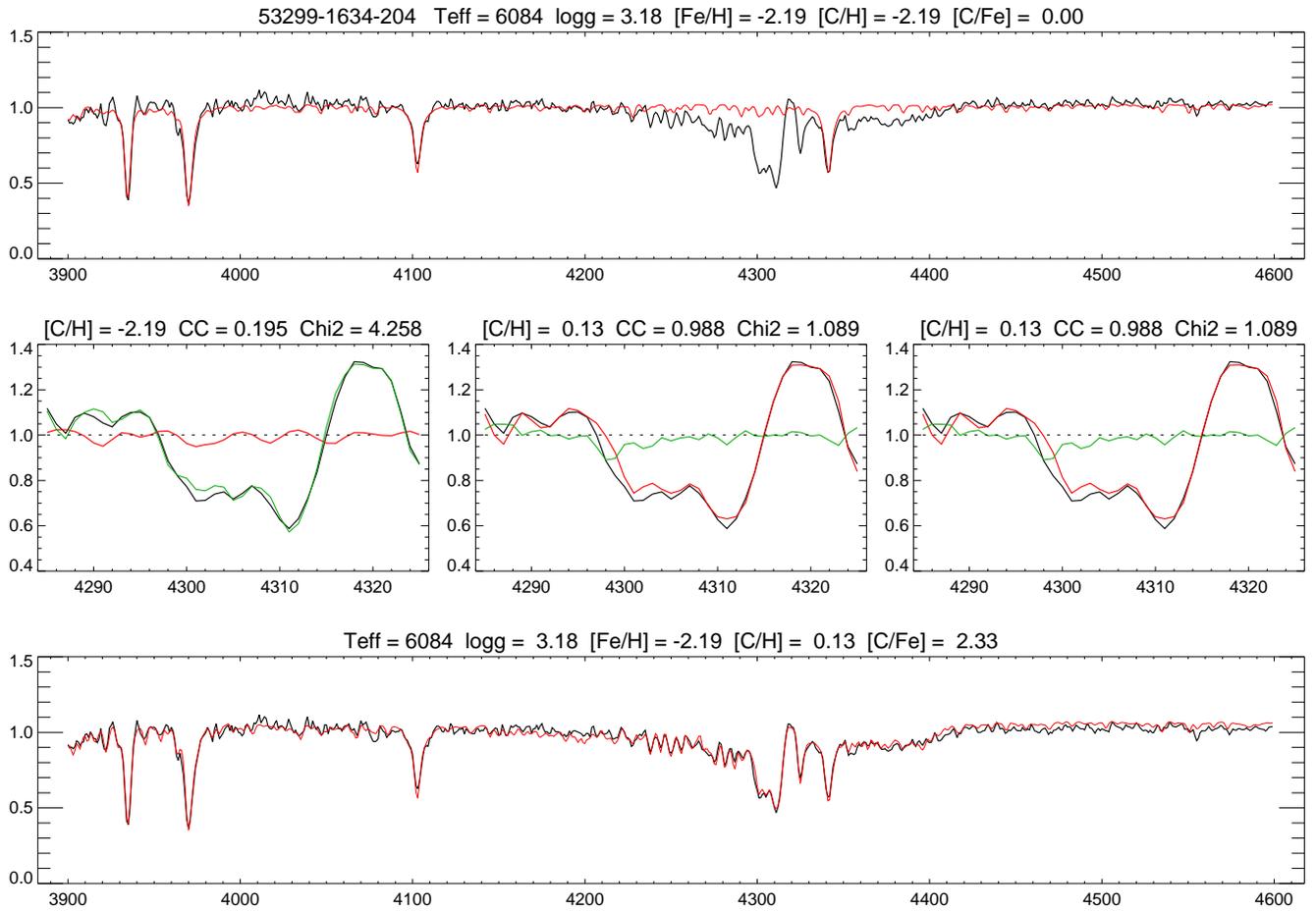}
\vspace{1cm}
\caption{Example of the carbon-to-iron abundance estimation procedure. Upper panel: Input
optical spectrum (black line) superposed with a synthetic spectrum (red line)
with [C/Fe] = 0 (the starting value). The legend at the top of this panel lists
the input values determined by the SSPP for \teff, \logg, and [Fe/H].
Middle panels:  Best matches to the CH G-band obtained from the three different search
ranges considered (left: [C/Fe] $< 0$; center:  [C/Fe] covering the full grid range;
right: [C/Fe] $> 0$). The red line in each case is the best match;
the green line is the ratio of the best match to the input spectrum.   The legend
above each of these panels provides the best estimate of [C/H], along with a
correlation coefficient between the synthetic spectrum and the input spectrum (CC),
and the reduced $\chi^2$ of the fit (Chi2). Note in the left-hand panel that no
value of [C/Fe] $< 0$ provides an acceptable match. The center and right-hand
panels shown are identical, because the routine converged to the same value for
the full grid search and the search restricted to [C/Fe] $> 0$. Lower panel:
Final adopted match, represented by the superposed red line. The legend above
this panel lists the resulting derived [C/Fe]. }
\end{center}
\end{figure}
\clearpage

\begin{figure}
\figurenum{2}
\begin{center}
\includegraphics[width=0.7\textwidth]{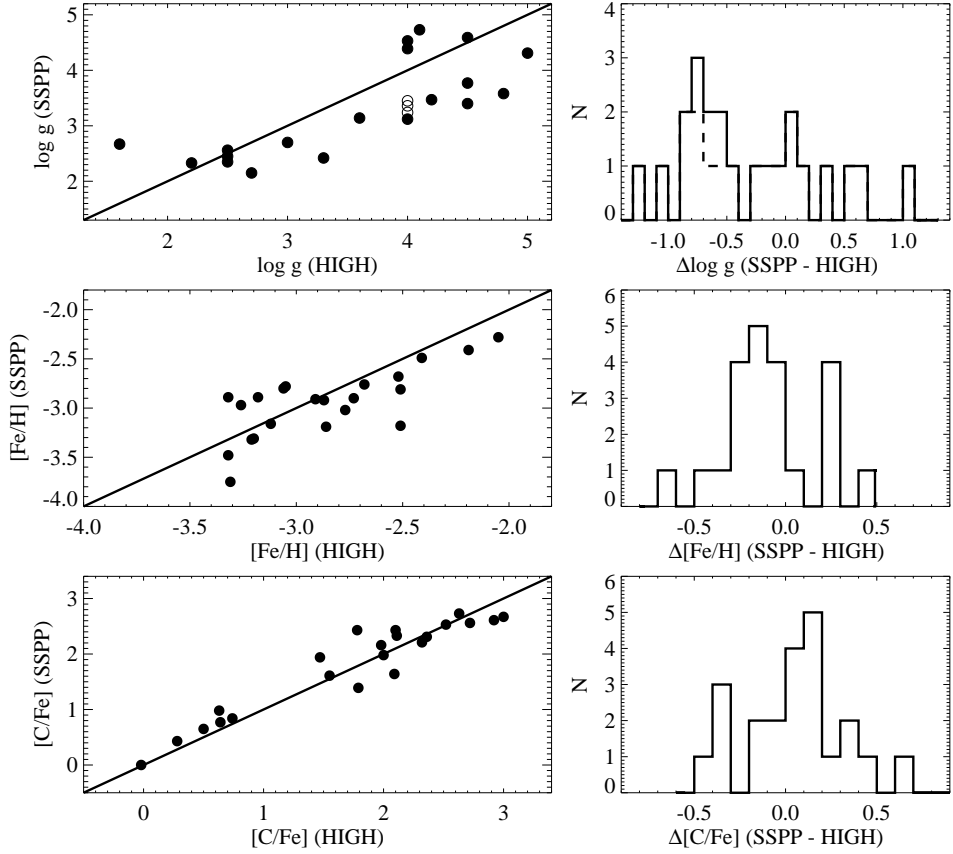}
\vspace{1cm}
\caption{Left panels: Comparison of the estimates of atmospheric parameters
(\logg\ , [Fe/H]), and carbonicity, [C/Fe], as derived from the low-resolution spectra (SSPP)
and the high-resolution follow-up spectra (HIGH). A one-to-one line is shown for
each. The open circles correspond to the stars with \logg\ values that were
assigned, than than derived (see text). Right panels: Histograms of the
differences between the SSSP and HIGH determinations. The dashed bins indicate
the stars with \logg\ values that were assigned, rather then derived. No
comparison is made for the \teff\ determinations, as the high-resolution
estimates for the majority of our comparison sample were taken directly from the
SSPP estimates (see text). Also see text for discussion of the reason for the
apparently large dispersion in determination of \logg\ .}
\end{center}
\end{figure}
\clearpage

\begin{figure}
\figurenum{3}
\hspace{-1cm}
\includegraphics[width=1.1\textwidth]{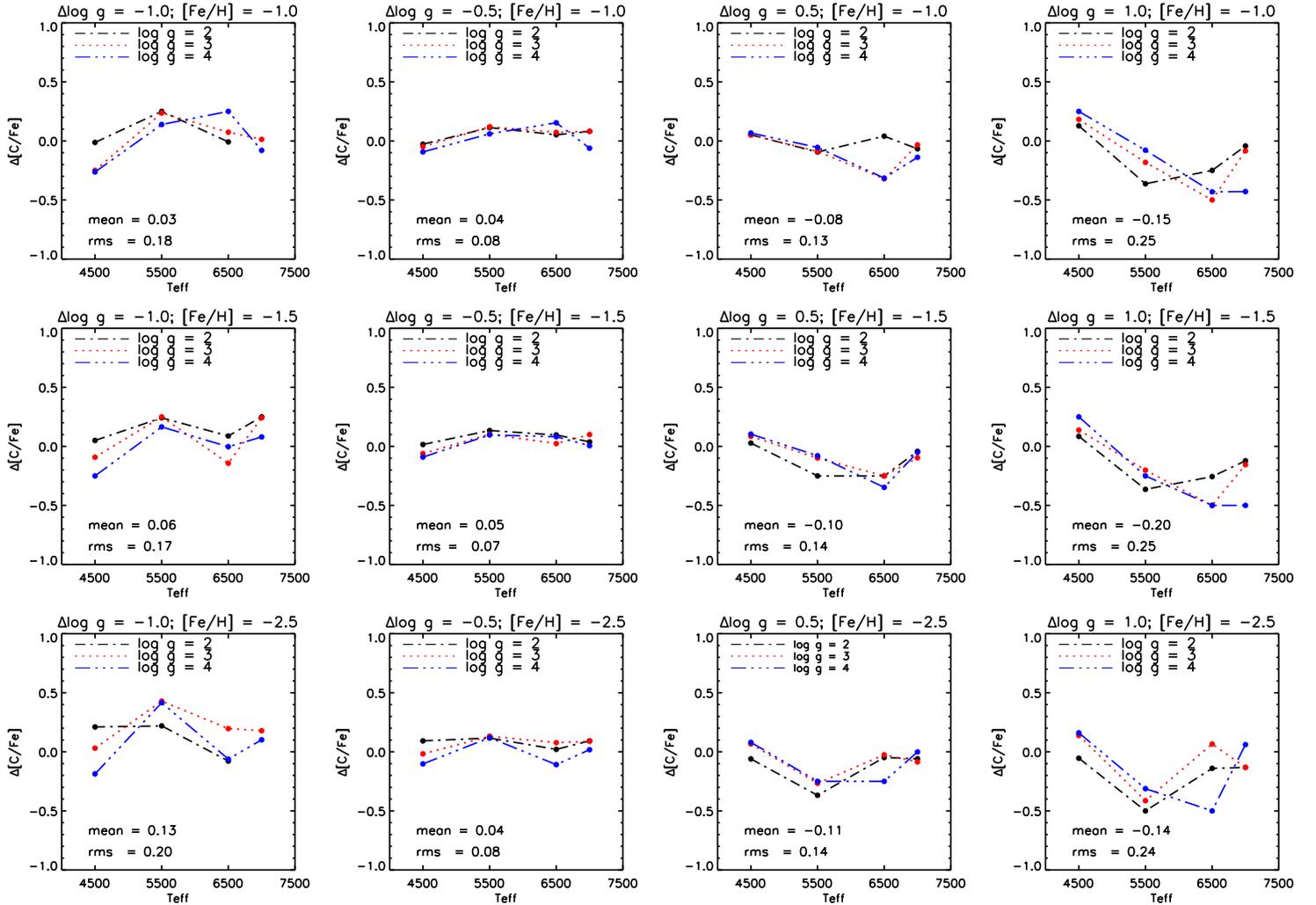}
\caption{Results of an experiment to estimate the impact of incorrect \logg\
estimates on the derivation of [C/Fe] (plotted as $\Delta$[C/Fe]), as described in
the text. The top row of panels correspond to synthetic spectra with [Fe/H] $=
-1.0$, the middle panels to [Fe/H] $= -1.5$, and the bottom panels to [Fe/H] $=
-2.5$. From left to right, the columns of panels correspond to input
perturbations in \logg\ of $-1.0$ dex, $-0.5$ dex, $+0.5$ dex, and $+1.0$ dex,
respectively.
The colored dots and lines shown in each panel corespond to input surface
gravities (prior to perturbing their values) of \logg\ = 2.0, 3.0, and
4.0. The mean zero-point offsets and rms variations in the derived [C/Fe],
relative to the known value across all \teff\ and \logg\ considered, are shown
for each panel. The case illustrated here is for [C/Fe] = 0.0; similar results
pertain to the case [C/Fe]$ = +1.5$.}
\end{figure}
\clearpage

\begin{figure}
\figurenum{4}
\begin{center}
\includegraphics[width=0.6\textwidth]{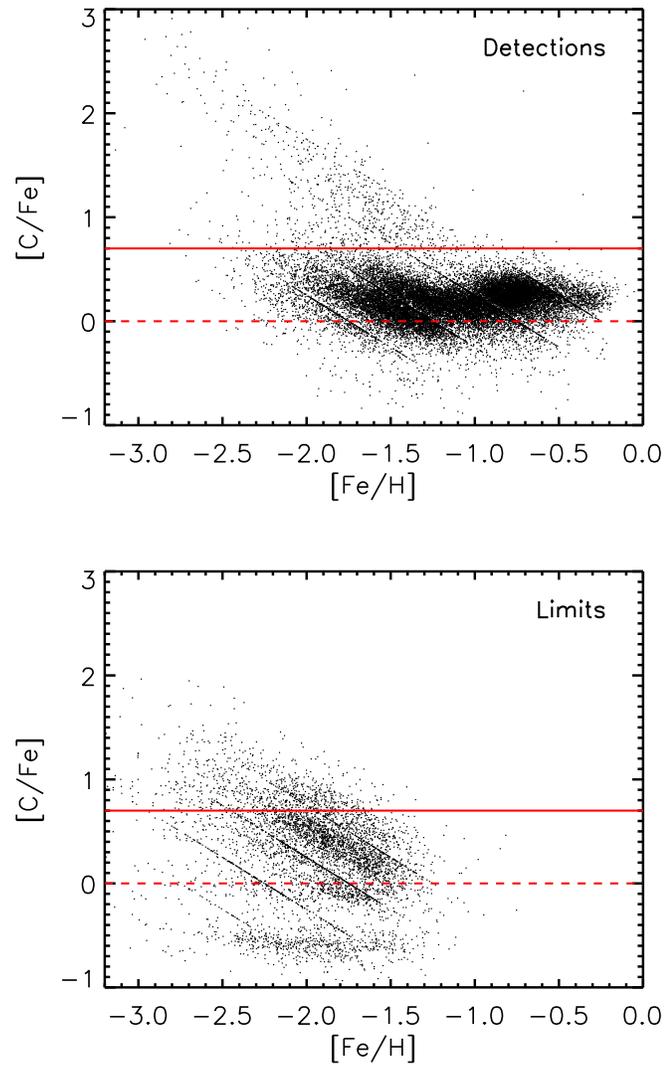}
\caption{Carbonicity, [C/Fe], as a function of metallicity,
[Fe/H], for subsample D (top panel) and subsample L (bottom panel). The dashed
red line represents the solar carbon abundance ratio [C/Fe] $= 0.0$; the
solid red line denotes the adopted limit that divides C-norm stars from C-rich
stars, [C/Fe] $= +0.7$. The ``ridge lines'' in both of these plots are due to grid
effects in the chi-square fitting procedure.}
\end{center}
\end{figure}
\clearpage

\begin{figure}
\figurenum{5}
\centering
\includegraphics[width=0.6\textwidth]{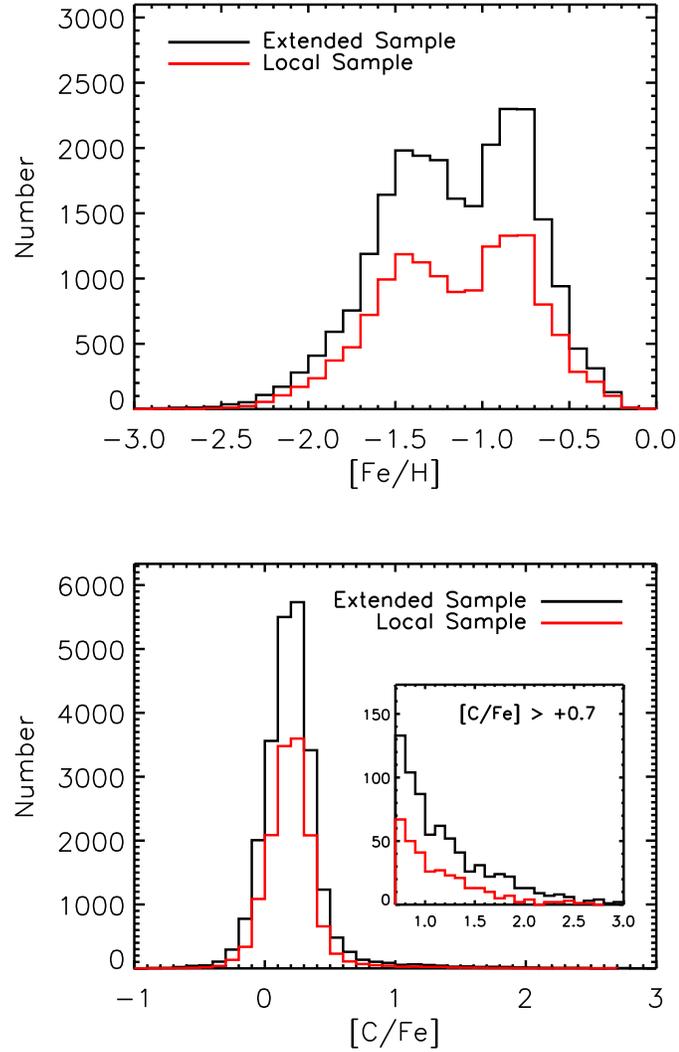}
\caption{Upper panel: As-observed metallicity distribution function (MDF)
for the Extended Sample of SDSS/SEGUE calibration stars (black histogram; $d <
10$ kpc, no restriction on $R$), and for the Local Sample (red histogram; $d <
4$ kpc; 7 kpc $ < R < 10$ kpc). Lower panel: As-observed carbonicity distribution function (CarDF)
for the Extended Sample of calibration stars (black histogram), and
for the Local Sample (red histogram). Note that the Extended Sample contains
significant numbers of dwarfs, main-sequence turnoff stars, and subgiants/giants (77\%,
13\%, and 10\%, respectively), while the Local Sample primarily comprises dwarfs
(88\%) and main-sequence turnoff stars (9\%), due to the larger volume explored
by the subgiants and giants. The inset shows a rescaled view of the
high-carbonicity tail.  For both panels, only stars with carbon detections are used.} 
\end{figure}
\clearpage

\begin{figure}
\figurenum{6}
\hspace{4.5cm}
\includegraphics[width=0.8\textwidth, angle=90]{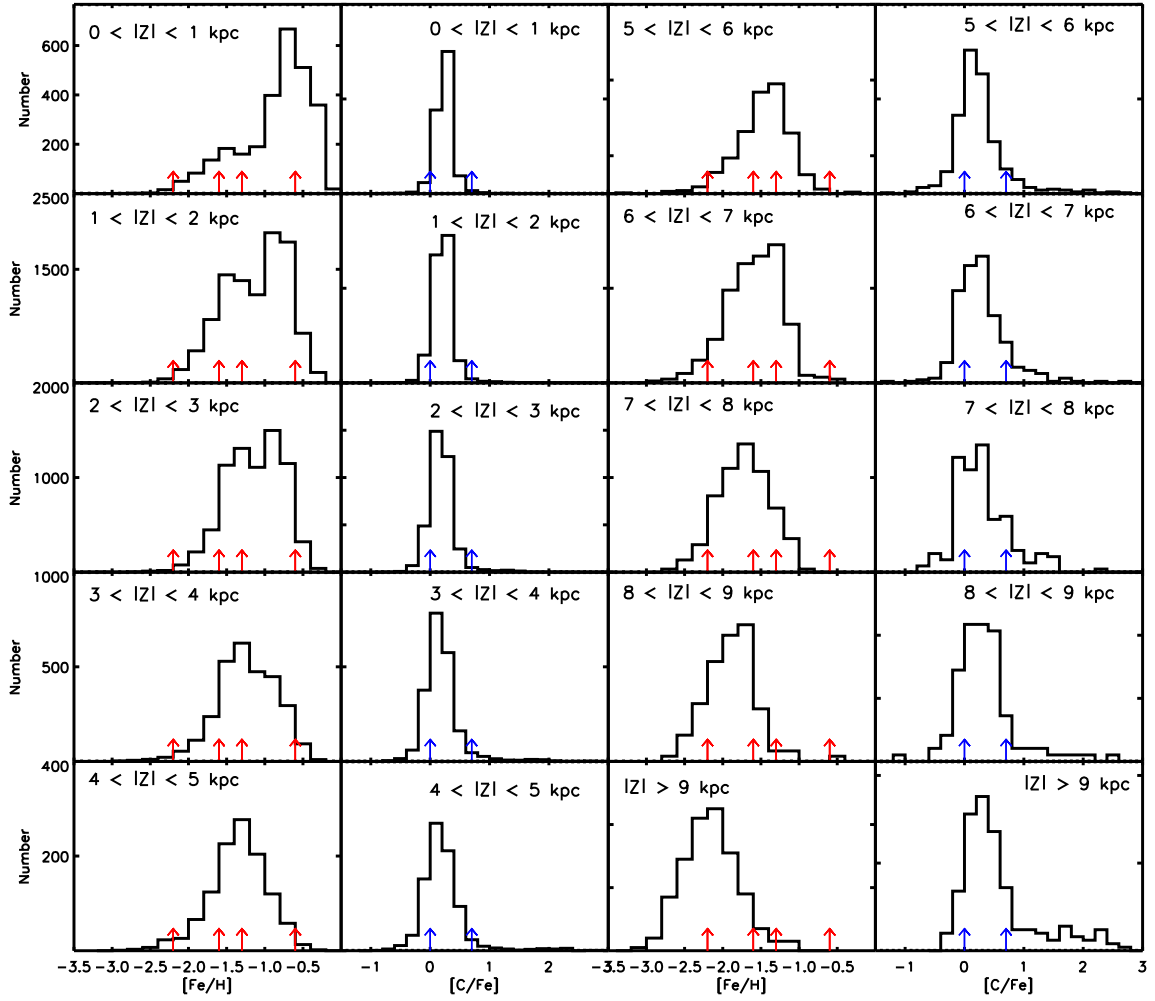}
\caption{First and third columns: As-observed metallicity distribution functions
(MDFs), for the Extended Sample of SDSS/SEGUE DR7 calibration stars, as a function
of vertical distance from the Galactic plane. The histograms represent the MDFs
obtained at different cuts of $|$Z$|$. The red arrows indicate the locations of
the metallicity peaks of the MDF for the thick disk ($-0.6$), the MWTD ($\sim
-1.3$), the inner halo ($-1.6$), and the outer halo ($-2.2$), respectively,
assigned by C10. Second and fourth columns: As-observed carbonicity distribution
functions (CarDFs), for the Extended Sample of SDSS/SEGUE DR7 calibration stars,
as a function of vertical distance from the Galactic plane. The blue arrows
show the location of the solar carbon-to-iron ratio ([C/Fe] = 0.0),
and the natural threshold that divides carbon-normal from carbon-rich stars
([C/Fe] $ = +0.7$). The histograms represent the CarDFs obtained at different cuts
of $|$Z$|$.  For all panels, only stars with carbon detections are used.}
\end{figure}

\begin{figure}
\figurenum{7}
\begin{center}
\includegraphics[width=0.9\textwidth]{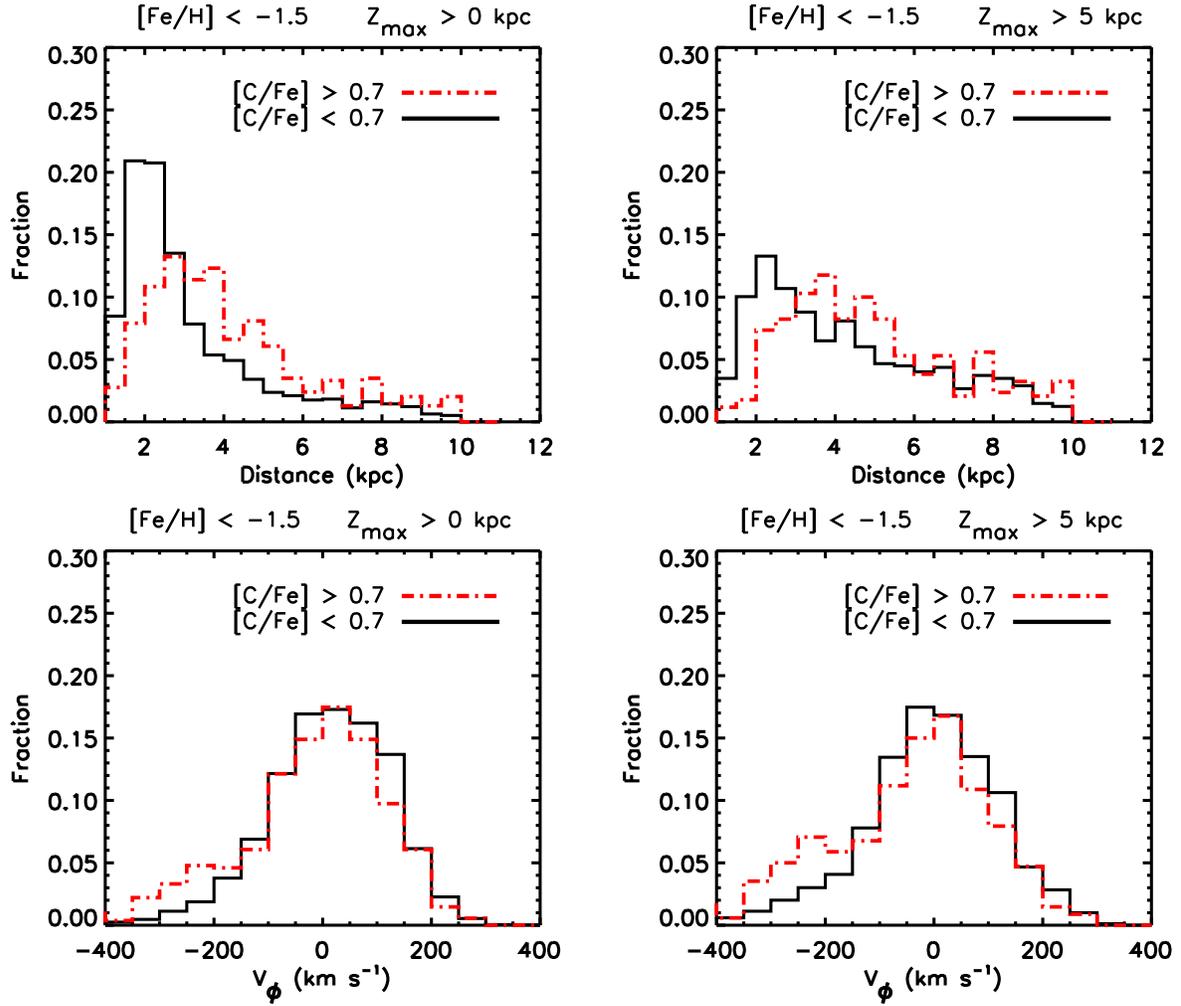}
\caption{Distributions of estimated distances (top panels) and rotational
velocity, V$_{\phi}$ (bottom panels), for stars considered CEMP ([C/Fe] $> +0.7$; dot-dashed red
histograms), and non-CEMP ([C/Fe] $< +0.7$; solid black histograms), for two cuts
on Z$_{\rm max}$. The stars are selected from the Extended Sample with [Fe/H] $<
-1.5$.  The left-hand column of panels includes stars at all Z$_{\rm max}$;
the right-hand column of panels is only for stars satisfying Z$_{\rm max}$ $>
5$ kpc.  For all panels, only stars with carbon detections are used.}
\end{center}
\end{figure}
\clearpage

\begin{figure}
\figurenum{8}
\begin{center}
\includegraphics[width=1.0\textwidth]{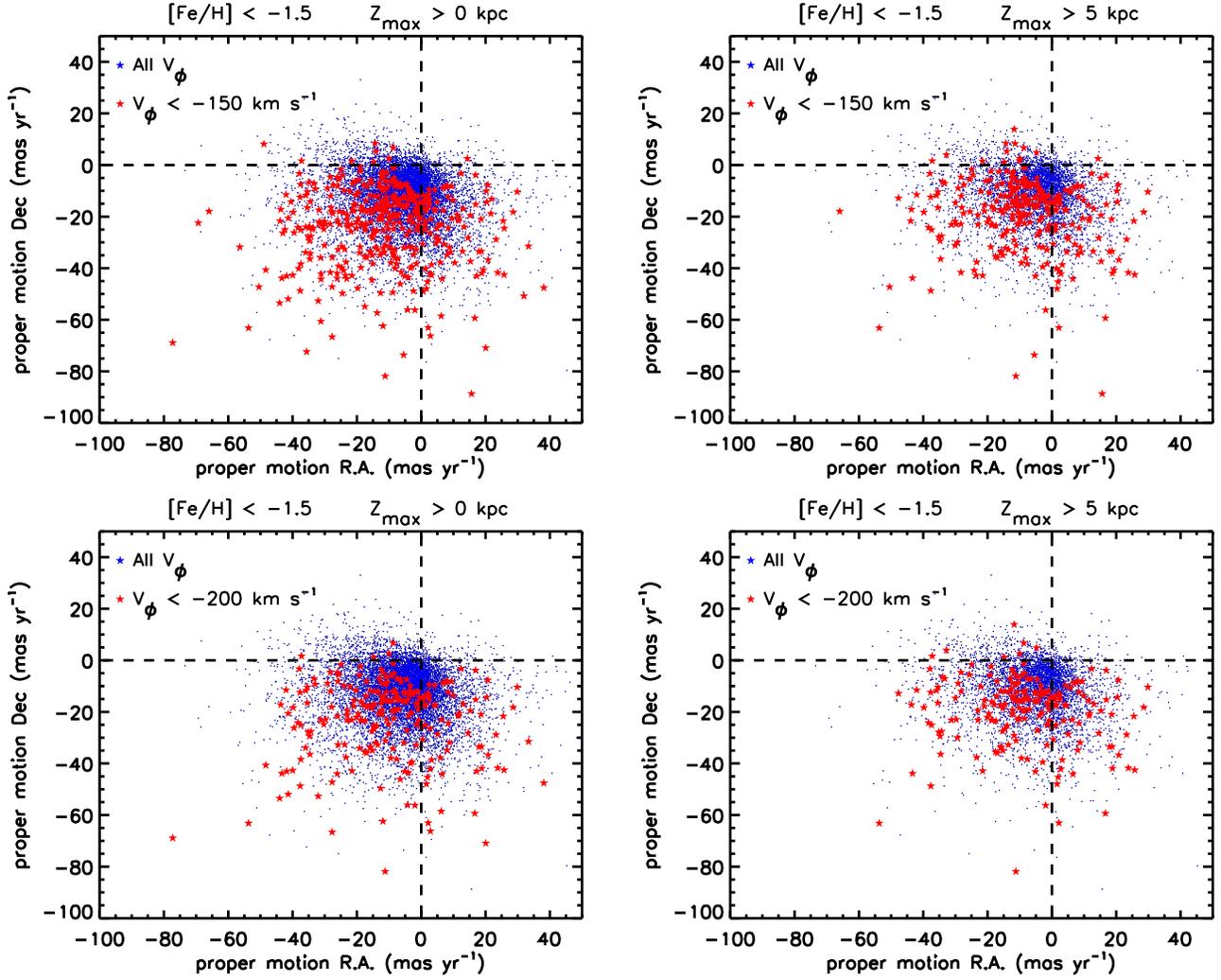}
\caption{Measured proper motion components for stars in the Extended Sample with [Fe/H] $< -1.5$
and two different cuts on \zmax\  . The left-hand column of panels includes stars at all Z$_{\rm max}$;
the right-hand column of panels is only for stars satisfying Z$_{\rm max}$ $>
5$ kpc.  The small blue dots represent the entire sample satisfying these cuts;
the filled red stars represent stars belonging to the highly
retrograde tails shown in the bottom panels of Figure 7.  The upper panels
correspond to the sample of retrograde stars with \vphi\ $< -150$ \kms\ ; the
lower panels correspond to the sample of retrograde stars with \vphi\ $< -200$
\kms\ .  Clear differences in the distributions of proper motions exist between
the stars in the highly retrograde tail and the rest of the sample.
The dashed lines provide a zero proper motion reference frame.
For all panels, only stars with carbon detections are used.}
\end{center}
\end{figure}
\clearpage

\begin{figure}
\centering
\figurenum{9}
\includegraphics[width=0.45\textwidth]{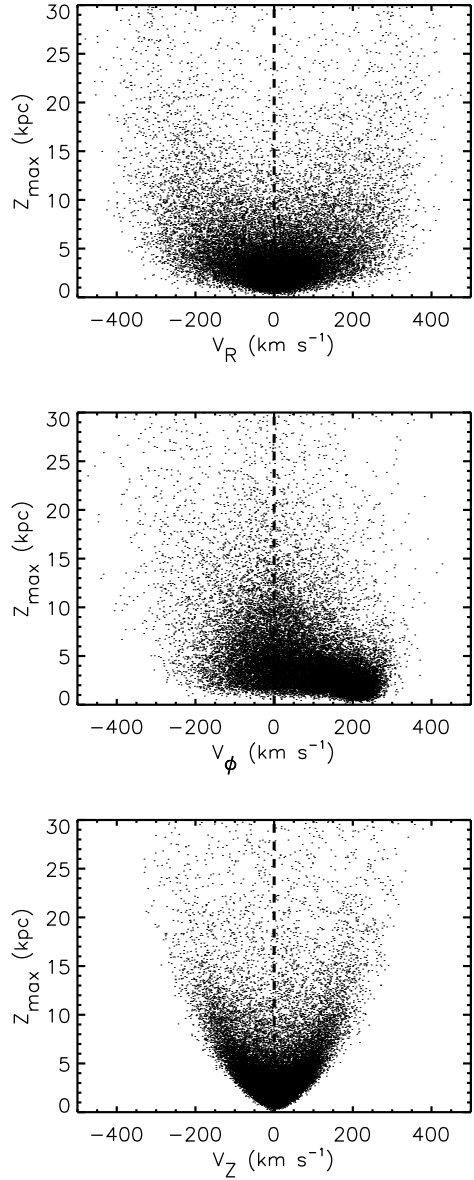}
\caption{The relationship between Z$_{\rm max}$ and the cylindrical velocity components
(V$_{R}$,V$_{\phi}$,V$_{Z}$) for the SDSS/SEGUE calibration stars in the
Extended Sample. The strong correlation between Z$_{\rm max}$ and the radial
velocity component shown in the upper panel, V$_{R}$, as well as with the
vertical velocity component shown in the lower panel, V$_{Z}$, is confirmed for
the Extended Sample. The middle panel exhibits no strong correlation between
Z$_{\rm max}$ and the rotational velocity component, V$_\phi$, other than that
expected from the presence of the thick-disk and halo populations. The dashed
line is a reference line at zero velocity for each component. Note, in the
middle panel, the clear excess of stars with retrograde motions for Z$_{\rm max}$
$>$ 10-15 kpc, which we associate with the outer-halo component. The panels include
all stars, regardless of their carbon status.}
\end{figure}
\clearpage

\begin{figure}
\figurenum{10}
\begin{center}
\includegraphics[width=0.5\textwidth]{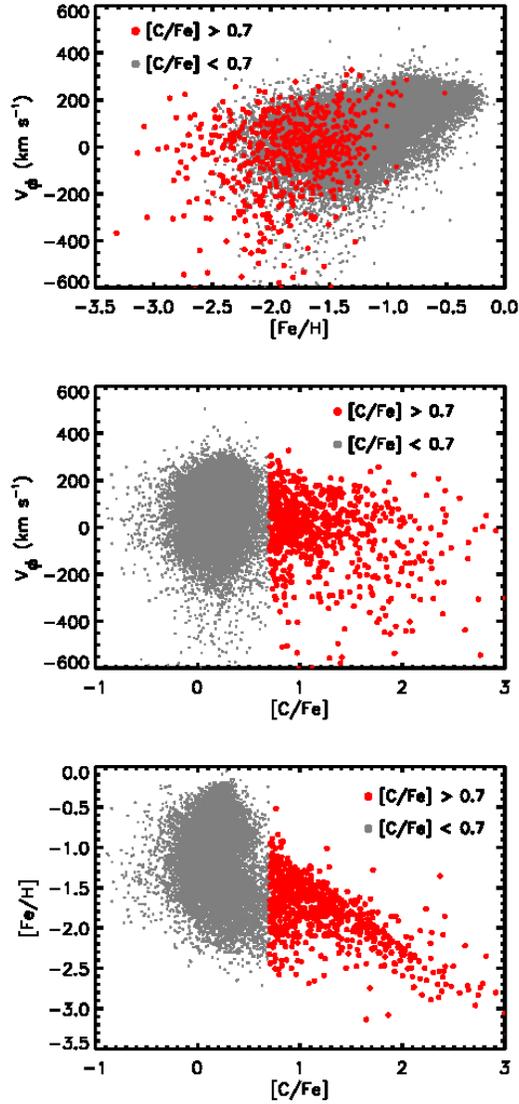}
\caption{Global behavior of the basic parameters, V$_{\phi}$, [Fe/H], and [C/Fe],
for the Extended Sample of SDSS/SEGUE calibration stars with detected CH
G-bands. Upper panel: Galactocentric rotational velocity as a function of the
metallicity. Middle panel: Galactocentric rotational velocity as a function of
[C/Fe]. Lower panel: Metallicity vs. carbonicity, [C/Fe]. In all panels, the gray dots
represent stars with low carbonicity, [C/Fe] $< +0.7$; the red dots
denote the stars with high carbonicity, [C/Fe] $>$ +0.7.}
\label{fig:windres}
\end{center}
\end{figure}
\clearpage

\begin{figure}
\figurenum{11}
\hspace{-1cm}
\vspace{2cm}
\includegraphics[width=1.1\textwidth]{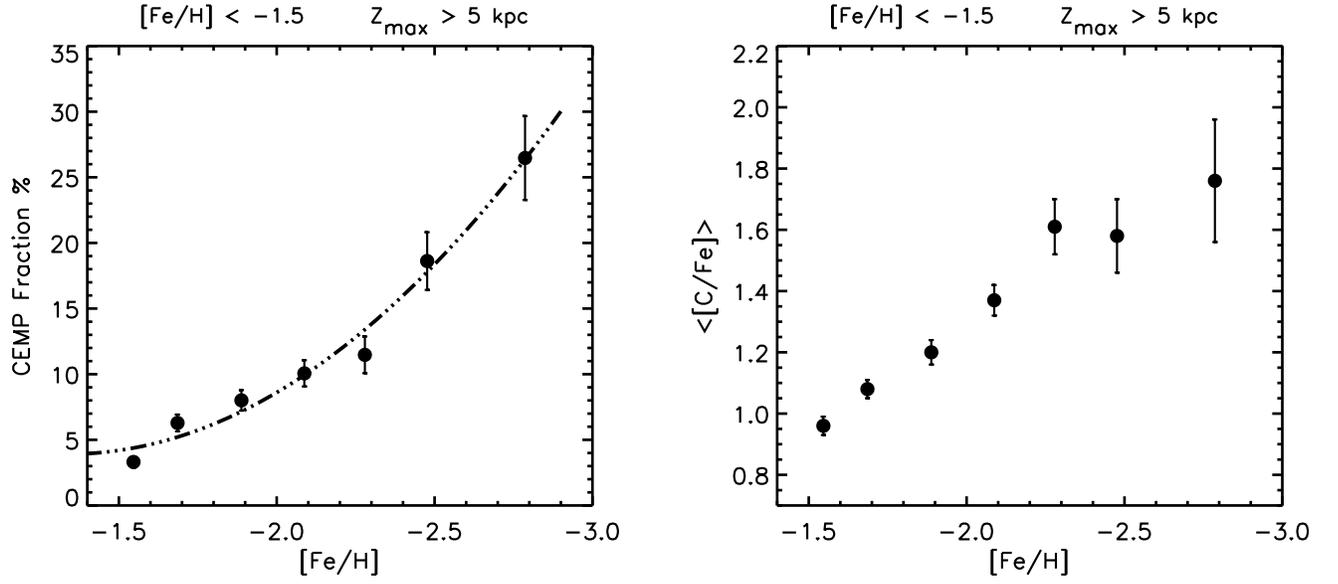}
\caption{Left panel: Global trend of the CEMP fraction, as a function of [Fe/H],
for the low-metallicity stars of the Extended Sample with \zmax\  $> 5$ kpc. Each bin
of metallicity has width $\Delta$[Fe/H] = 0.2 dex, with the exception of the
lowest-metallicity bin, which includes all stars with [Fe/H] $< -2.6$. The error
bars are evaluated with the jackknife method. The calculation of CEMP star
frequency is made using Eqn. 2, which includes all stars of
known carbon status, including the L subsample.  A second-order polynomial fit
to the observed distribution is shown by the dot-dashed line.
Right panel: Global trend of the mean carbonicity, $\langle$[C/Fe]$\rangle$, as
a function of [Fe/H]. Only those stars with detected CH G-bands are used. The
bins are the same as used in the left panel. Errors are the standard error in
the mean.}
\end{figure}
\clearpage

\begin{figure}
\figurenum{12}
\begin{center}
\includegraphics[width=0.8\textwidth]{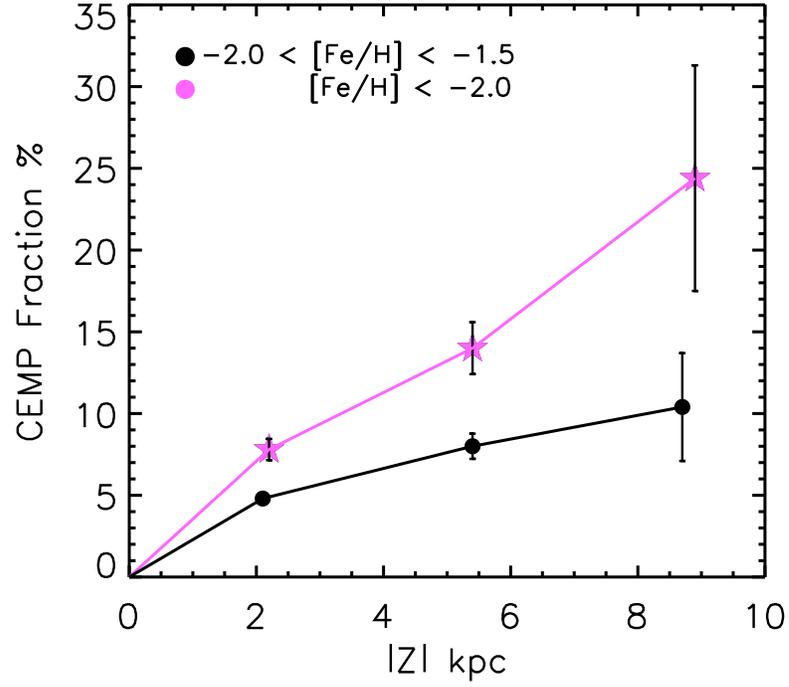}
\caption{Global trend of the
CEMP star fraction, as a function of vertical distance from the Galactic plane, $|$Z$|$,
for the low-metallicity stars of the Extended Sample. The stars with $-$2.0 $<$[Fe/H] $<
-1.5$ are shown as filled black circles; those with [Fe/H] $< -2.0$ are shown
as magenta stars.  Each bin has a width of $\Delta$$|$Z$|$ = 4 kpc, with the
exception of the last bin, which cuts off at 10 kpc, the limiting distance
of the Extended Sample. Errors are derived using the jackknife approach.  The calculation
of CEMP star frequency is made using Eqn. 2, which includes all stars of
known carbon status, including the L subsample.  Note the clear dependence of
CEMP fraction on height above the Galactic plane, for both metallicity regimes.}
\end{center}
\end{figure}
\clearpage

\begin{figure}
\figurenum{13}
\begin{center}
\includegraphics[width=0.45\textwidth]{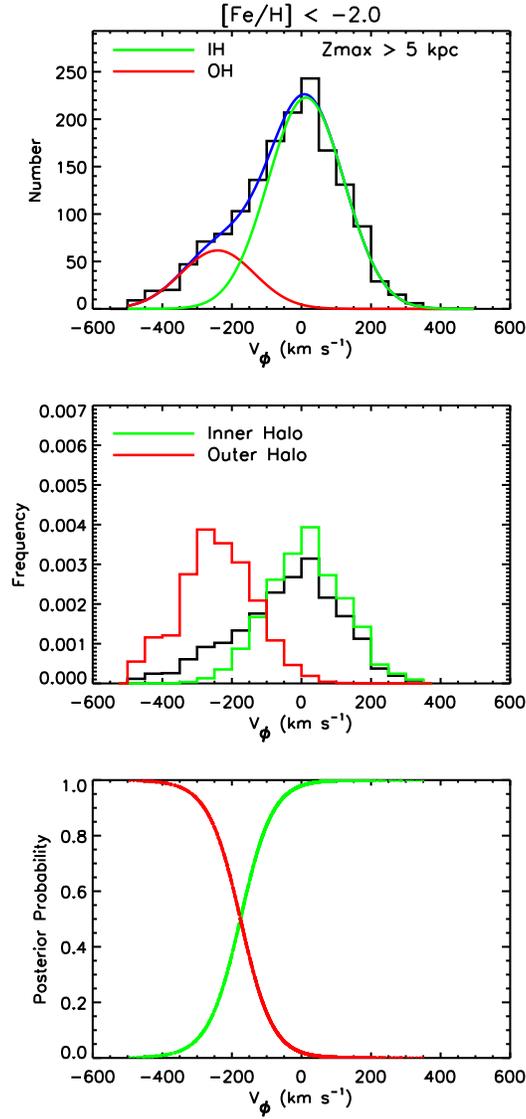}
\caption{Rotational properties for the Extended Sample of SDSS/SEGUE DR7 calibration
stars with metallicity [Fe/H] $<$ $-$2.0 and Z$_{\rm max}$ $>$ 5 kpc.
Upper panel: The black histogram represents the observed
distribution of V$_{\phi}$; the green (inner halo), and red (outer halo)
curves show the results of the Extreme Deconvolution analysis. Middle panel: The
observed velocity distribution function (black histogram), and the inner- and
outer-halo weighted velocity distribution functions (frequencies), denoted by the green and
the red histograms, respectively. In this panel, all the distributions are
normalized and take into account the membership probability of each star.
Lower panel: The posterior probability for the inner- and outer-halo components as a
function of the rotational velocity, V$_{\phi}$.  Note that these plots use all
stars in the Extended Sample satisfying the stated cuts, regardless of whether
or not they have detected CH G-bands.}
\end{center}
\end{figure}
\clearpage

\begin{figure}
\figurenum{14}
\centering
\includegraphics[width=0.6\textwidth]{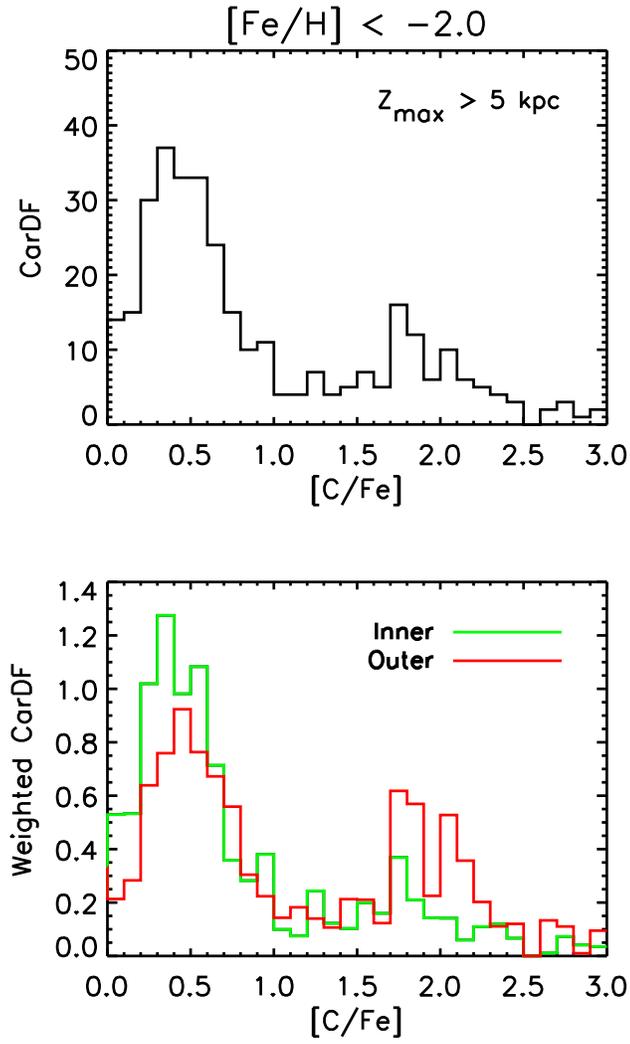}
\caption{The carbonicity distribution functions (CarDF) for the Extended Sample
with [Fe/H] $<$ $-$2.0 and Z$_{\rm max}$ $>$ 5 kpc and with detected CH G-bands.
Top panel: CarDF for the entire sample.  Bottom panel: Weighted CarDF
for the inner- and outer-halo components, shown by the green and red histograms,
respectively.}
\end{figure}
\clearpage

\begin{figure}
\figurenum{15}
\hspace{-2cm}
\includegraphics[width=1.2\textwidth]{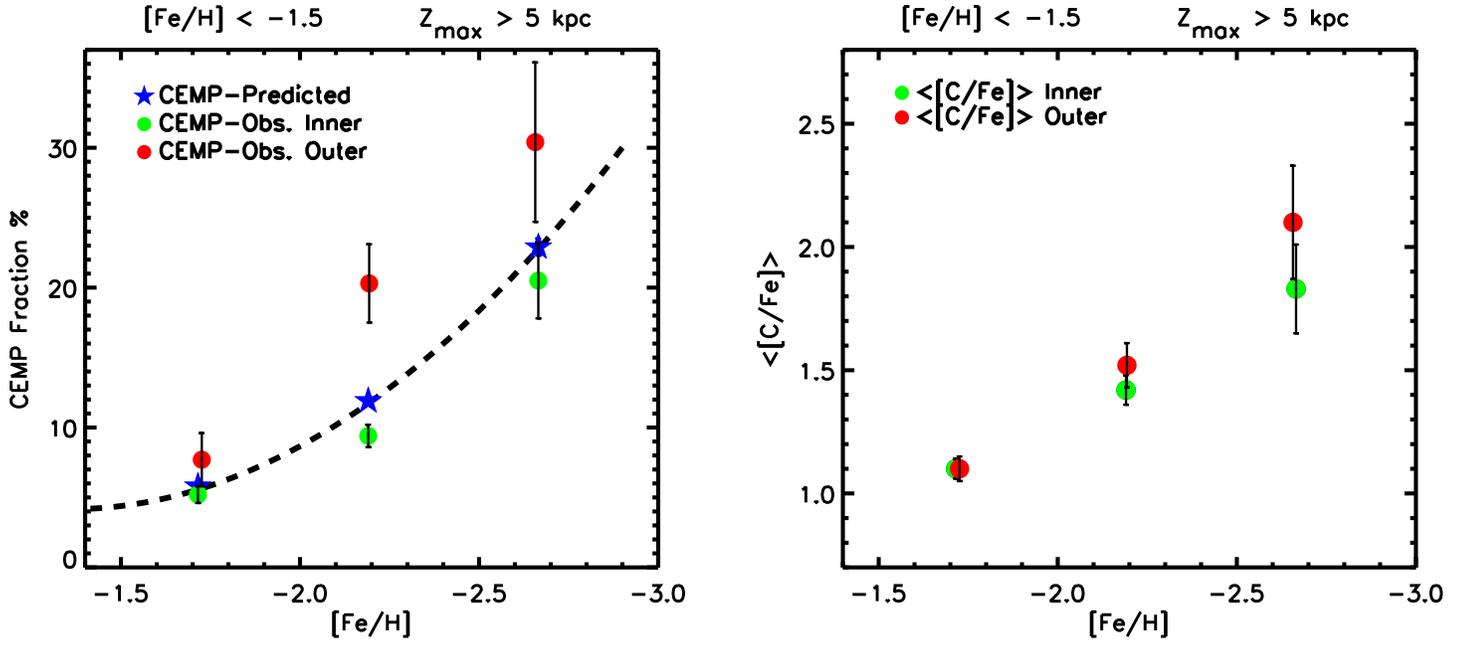}

\caption{Left panel: Global trend of the CEMP star fraction as a function
of [Fe/H] for the low-metallicity stars of the Extended Sample with
\zmax\  $> 5$ kpc.  The dashed curve is a second-order polynomial fit representing
the global trend. The blue filled stars represent the predicted values of the
CEMP fractions in each bin of metallicity, having a width of $\Delta$[Fe/H]
= 0.5 dex. The green and red filled circles show the location of the observed
CEMP star fractions for the inner halo and outer halo, respectively, based on
the XD method with the hard-cut-in-probability approach (see text). The error
bars are evaluated with the jackknife method. The calculation is made using Eqn.
2, which includes all stars of known carbon status, including the L subsample.
Right panel: Global trend of the mean carbonicity, $\langle$[C/Fe]$\rangle$, as
a function of [Fe/H]. Only those stars with detected CH G-bands are used. The
green and red filled circles show the mean carbonicity for the inner halo and
outer halo, respectively, based on the Extreme Deconvolution method with the
hard-cut-in-probability approach (see text). The bins are the same as used in
the left panel. Errors are the standard error in the mean. No significant
difference is seen for the inner- and outer-halo subsamples.}
\end{figure}
\clearpage

\end{document}